\documentclass[11pt,a4paper]{article}

\usepackage{enumitem}

\usepackage{amsmath,amscd}
\usepackage{amsfonts}
\usepackage{amssymb,scalerel,stackengine}
\usepackage{amsmath,amsfonts,amssymb,amsthm,bbm,bm}
\usepackage{relsize}
\usepackage{empheq}
\usepackage{graphicx,shortvrb}
\usepackage{comment}
\usepackage{cite}
\usepackage[utf8]{inputenc}
\usepackage[colorlinks=true,linkcolor=blue, linktoc=page, urlcolor=blue,citecolor=magenta]{hyperref}

\numberwithin{equation}{section}

\hypersetup{
    colorlinks,%
    citecolor=blue,%
    filecolor=blue,%
    linkcolor=blue,%
   urlcolor=blue,
   linktoc=page
}

\def\mn{{\mu\nu}}
\def\de{\delta}

\def\pd{\partial}
\def\p{\partial}

\def\eps{\epsilon}

\def\bK{\boldsymbol{K}}
\def\bA{\boldsymbol{A}}
\def\bB{\boldsymbol{B}}
\def\bC{\boldsymbol{C}}

\def\bS{\boldsymbol{S}}
\def\bU{\boldsymbol{U}}
\def\bV{\boldsymbol{V}}

\newcommand*\widefbox[1]{\fbox{\hspace{2em}#1\hspace{2em}}}

\newcommand\widecheck[1]{%
\savestack{\tmpbox}{\stretchto{%
  \scaleto{%
    \scalerel*[\widthof{\ensuremath{#1}}]{\kern-.6pt\bigwedge\kern-.6pt}%
    {\rule[-\textheight/2]{1ex}{\textheight}}
  }{\textheight}%
}{0.5ex}}%
\stackon[1pt]{#1}{\scalebox{-1}{\tmpbox}}%
}

\newcommand{\bea}{\begin{eqnarray}}
\newcommand{\eea}{\end{eqnarray}}

\newcommand{\geo}[1]{{ \textcolor{blue}{{#1}}}}

\begin{document}

\setcounter{tocdepth}{2}

\begin{titlepage}

\begin{flushright}\vspace{-3cm}
{\small
\today }\end{flushright}
\vspace{0.5cm}

\begin{center}

{{ \Large{\bf{The Poincar\'e and BMS flux-balance laws \\ \vspace{8pt} {with  application to binary systems}}}}}
\vspace{5mm}

\bigskip

\centerline{\large{\bf{Geoffrey Comp\`{e}re$^\dagger$\footnote{email: gcompere@ulb.ac.be}, Roberto Oliveri$^*$\footnote{email: roliveri@fzu.cz} and Ali Seraj$^\dagger$\footnote{email: aseraj@ulb.ac.be}}}}\vspace{2pt}

\vspace{2mm}
\normalsize
\bigskip\medskip
\textit{{}$^\dagger$ Centre for Gravitational Waves, Universit\'{e} Libre de Bruxelles, \\
 International Solvay Institutes, CP 231, B-1050 Brussels, Belgium}\\
\textit{{}$^*$ CEICO, Institute of Physics of the Czech Academy of Sciences,\\
Na Slovance 2, 182 21 Praha 8, Czech Republic }

\vspace{15mm}

\begin{abstract}
\noindent

{Asymptotically flat spacetimes admit both supertranslations and Lorentz transformations as asymptotic symmetries. Furthermore, they admit super-Lorentz transformations, namely superrotations and superboosts, as outer symmetries associated with super-angular momentum and super-center-of-mass charges. In this paper, we present  comprehensively the flux-balance laws for all such BMS charges. We distinguish the Poincar\'e flux-balance laws from the proper BMS flux-balance laws associated with the three relevant memory effects defined from the shear, namely, the displacement, spin and center-of-mass memory effects. We scrutinize the prescriptions used to define the angular momentum and center-of-mass. In addition, we provide the exact form of all Poincar\'e and proper BMS flux-balance laws in terms of radiative symmetric tracefree multipoles. Fluxes of energy, angular momentum and octupole super-angular momentum arise at 2.5PN, fluxes of quadrupole supermomentum arise at 3PN and fluxes of momentum, center-of-mass and octupole super-center-of-mass arise at 3.5PN. We also show that the BMS flux-balance laws lead to integro-differential consistency constraints on the radiation-reaction forces acting on the sources. Finally, we derive the exact form of all BMS charges for both an initial Kerr binary and a final Kerr black hole in an arbitrary Lorentz and supertranslation frame, which allows to derive exact constraints on gravitational waveforms produced by binary black hole mergers from each BMS flux-balance law.  
}
\end{abstract}

\end{center}

\end{titlepage}

\tableofcontents

\setcounter{footnote}{0}

\section{Introduction and outline}

The study of gravitational radiation has a long history enlightened by the seminal work of Bondi, van der Burg, Metzner and Sachs \cite{1962RSPSA.269...21B,1962RSPSA.270..103S}. There, it was realized that asymptotically flat spacetimes not only lead to Poincar\'e charges but also to so-called supermomenta associated with supertranslation asymptotic symmetries. Following the pioneer work of \cite{Barnich:2009se,Barnich:2011ty,Barnich:2011mi}, it was subsequently realized that super-Lorentz charges, associated with the so-called superrotations and superboosts, are also finite surface charges for standard asymptotically flat spacetimes \cite{Flanagan:2015pxa,Compere:2016hzt,Hawking:2016sgy}, even though the corresponding diffeomorphisms are not asymptotic symmetries but external symmetries \cite{Compere:2016jwb}.\footnote{Alternatively, either asymptotic Virasoro$\times$Virasoro \cite{deBoer:2003vf,Barnich:2009se,Barnich:2010eb,Barnich:2011mi} or smooth Diff$(S^2)$ symmetries \cite{Campiglia:2014yka,Campiglia:2015yka} can be considered to act on an extended notion of asymptotically flat spacetimes that encompasses cosmic events such as cosmic string decays or Robinson-Trautman waves \cite{Podolsky:2002sa,Podolsky:2010xh,Podolsky:2016mqg,Griffiths:2002hj,Griffiths:2002gm,Barnich:2016lyg,Strominger:2016wns,Compere:2018ylh,Flanagan:2019vbl,Bhattacharjee:2019jaf}. In this paper, we only consider standard asymptotically flat spacetimes, \emph{i.e.}, that can be written as $g_{\mu\nu} = \eta_{\mu\nu}+ O(1/r)$ for $r \rightarrow \infty$.} 
Each such ``extended BMS'' charge is associated with a flux-balance law that relates all forms of radiation reaching null infinity (including matter radiation) to the difference of BMS surface charges at initial and final retarded times \cite{Barnich:2011ty,Barnich:2011mi,Barnich:2013axa,Flanagan:2015pxa,Compere:2016hzt,Hawking:2016sgy}. The extended BMS flux-balance laws are complete at second order in the large radius expansion, in the sense that no other evolution equation arises from Einstein's equations in Bondi gauge at that order, though an infinite tower  of subleading flux-balance laws are implied by Einstein's equations \cite{vdBurg,Barnich:2010eb}. While such leading and subleading flux-balance laws have already appeared in the literature \cite{Payne:1982aa,Barnich:2010eb,Barnich:2011ty,Barnich:2011mi,Barnich:2013axa,Strominger:2013jfa,Strominger:2014pwa,Flanagan:2015pxa,Barnich:2016lyg,Compere:2018ylh,Nichols:2017rqr,Nichols:2018qac,Bonga:2018gzr,Distler:2018rwu,Ashtekar:2019viz,Ashtekar:2019rpv}, no unified presentation has yet been given that provides a comprehensive first-principle derivation of all BMS laws for standard asymptotically flat spacetimes which includes their explicit relationship to memory effects. The first objective of this paper is to provide such a unified presentation, complementing the remarkable work of Nichols \cite{Nichols:2017rqr,Nichols:2018qac}. The ten Poincar\'e flux-balance laws will be uniquely identified after providing the prescription for the angular momentum matching with \cite{Thorne:1980ru,Ashtekar:1981bq,Dray:1984rfa,Wald:1999wa,Barnich:2010eb,Barnich:2011mi,Distler:2018rwu,Compere:2018ylh} and for the center-of-mass matching with \cite{Kozameh:2013bha}. The proper BMS flux-balance laws will be explicitly related to the three relevant memories defined from the Bondi shear, namely the displacement memory \cite{Zeldovich:1974aa,1977ApJ...216..610T,1978ApJ...220.1107T,1978Natur.274..565T,1978ApJ...224...62K,1985ZhETF..89..744B,1987Natur.327..123B,Blanchet:1987wq,Christodoulou:1991cr,PhysRevD.44.R2945,Blanchet:1992br,0264-9381-9-6-018,Favata:2008yd,Pollney:2010hs,Wang:2014zls,Arzoumanian:2015cxr,Lasky:2016knh,Madler:2016ggp,Hubner:2019sly}, the spin memory \cite{Arun:2004ff,Pasterski:2015tva,Nichols:2017rqr,Flanagan:2018yzh,Himwich:2019qmj} and the center-of-mass memory \cite{Nichols:2018qac}. 

The BMS flux-balance laws provide explicit consistency constraints on waveforms generated by compact binary mergers that are derived within General Relativity. Such constraints have already been explicitly discussed for momenta and BMS supermomenta \cite{1962RSPSA.269...21B,1962RSPSA.270..103S}. The energy-momentum flux-balance laws allow to deduce the final mass and velocity kick after the merger as a function of the initial parameters of the binary. The proper supermomentum flux-balance laws instead allow to deduce the total displacement memory as a function of the radiation and initial parameters of the merger \cite{Payne:1982aa,1987Natur.327..123B,Blanchet:1987wq,Christodoulou:1991cr,PhysRevD.44.R2945,Blanchet:1992br}. The Lorentz flux-balance laws similarly allow to deduce the final angular momentum and center-of-mass shift after the merger, though the choice of the supertranslation and Lorentz frame introduces complications  \cite{Ashtekar:2019rpv}. Finally, the super-Lorentz flux-balance laws allow to deduce the spin and center-of-mass memories \cite{Nichols:2017rqr,Nichols:2018qac}. Our second objective is to provide an explicit formulation of all BMS consistency constraints that could be directly used by numerical relativists and gravitational wave data analysts. We will achieve this goal by providing a unified presentation of all BMS global constraints and, in addition, by providing the explicit BMS charges of the initial and final states of black hole binary mergers, which completes partial results previously obtained in the literature \cite{1962RSPSA.269...21B,1962RSPSA.270..103S,Strominger:2013jfa,Blanchet:2004re,Flanagan:2015pxa,Compere:2016jwb,Compere:2016hzt,Madler:2018tkl,Ashtekar:2019rpv}.

A crucial analytic method used to derive gravitational waveforms for binary mergers is the post-Newtonian/post-Minkowskian matched asymptotic expansion scheme between the source near-zone and the radiative far-zone; see \emph{e.g.}, \cite{1976ApJ...210..764W,Thorne:1980ru,Blanchet:1985sp,Will:1996zj} as well as the effective field theory methods \cite{Goldberger:2004jt,Porto:2016pyg}. The BMS formalism allows for an exact solution of the far-zone gravitational field. The third objective of this paper is to exploit this far-zone solution to derive the exact form of all Poincar\'e and proper BMS flux-balance laws in the symmetric trace-free (STF) radiative multipole expansion to all orders in the Newton's gravitational constant $G$ and the speed of light $c$. This completes the existing results in the literature \cite{Thorne:1980ru,Ruiz:2007yx,Nichols:2017rqr,Nichols:2018qac,Blanchet:2018yqa} and thereby constitutes a resolved sub-problem of the post-Newtonian/post-Minkowskian matched asymptotic expansion scheme. In turn, these multipolar BMS flux-balance laws can be expressed in terms of source parameters. We will elaborate that -- in the case of a compact binary inspiral -- the BMS flux-balance laws lead to towers of coupled integro-differential constraints on the source parameters.

The rest of the paper is organized as follows. After a short review of the BMS formalism in Section \ref{shortrev}, we will present the complete set of BMS flux-balance laws in simplified form in Section \ref{sec:BMS}. In particular, we will derive the angular momentum and center-of-mass flux-balance laws and extensively discuss their comparison with the literature. Section \ref{sec:flux} is devoted to the multipole expansion of the BMS flux-balance laws. We present the explicit constraints resulting from the time integrated  BMS flux-balance laws for binary black hole mergers in Section \ref{sec:glob}. We conclude in Section \ref{sec:ccl}. The derivation of the Kerr black hole in an arbitrary supertranslation and Lorentz frame in relegated to Appendix \ref{app:Kerr}, while Appendices \ref{App:T1T2}, \ref{app:int} and \ref{app:sph} provide technical details on the multipole expansion and on the integration of STF tensors and a comparison between STF and spherical harmonics.

\paragraph{Notations and conventions}
In this paper, we explicitly keep track of the speed of light $c$ and of Newton's gravitational constant $G$ in all formulae. Upper-case Latin letters $\{A, B, C, \dots \}$ label indices of tensors defined over the sphere; lower-case Latin letters $\{i, j, k, \dots \}$ label indices of tensors defined over the unit sphere embedded into Euclidean space, with unit normal vector $\vec{n}=(\sin\theta \cos\phi,\sin\theta \sin\phi,\cos\theta)$. The Einstein summation convention will always be implicit over repeated indices, even if both are raised or lowered. The unit normalized integral of a function over the sphere, $\int_S \frac{d^2\Omega}{4\pi} f(x^A)$, will be denoted by the shorthand notation $\oint_S f(x^A)$. An overdot is used to denote derivation with respect to retarded time, while the $k$-th retarded time derivative of $f$ is denoted as $\stackrel{(k)}{f}$.

\section{Brief review of the BMS formalism}
\label{shortrev}

The study of gravitational radiation close to future null infinity $\mathcal{I}^+$ can be suitably studied in Bondi gauge  \cite{1962RSPSA.269...21B,1962RSPSA.270..103S}. This section is aimed at reviewing the main ingredients of the BMS formalism that we will need and at spelling our conventions. For more extended reviews, see \emph{e.g.},  \cite{Barnich:2010eb,Flanagan:2015pxa,Madler:2016xju,Hawking:2016sgy,Strominger:2017zoo,Ashtekar:2018lor,Compere:2018ylh,Compere:2019qed}.

\subsection{Metric in Bondi gauge}

We use retarded coordinates $(u,r,x^A)$ near future null infinity. Bondi gauge is defined from the gauge fixing conditions $g_{rr}=g_{rA}=0$ and $\p_r \det \left(r^{-2}g_{AB}\right) = 0$, the latter fixing the radial coordinate $r$ to be the luminosity (areal) distance. We consider General Relativity coupled to  a matter stress-tensor obeying the following asymptotic conditions as $r \rightarrow \infty$ (as  in \cite{Flanagan:2015pxa} which slightly generalizes \cite{Strominger:2014pwa}) 
\begin{subequations}
\begin{align}
T_{uu} &= r^{-2}\hat T_{uu}(u,x^A) + O(r^{-3}), \qquad &T_{ur} &= O(r^{-4}), \\
T_{rr} &= r^{-4}\hat T_{rr}(u,x^A) + O(r^{-5}),\qquad &T_{uA} &= r^{-2}\hat T_{uA}(u,x^A) + O(r^{-3}),  \\
T_{rA} &= r^{-3} \hat T_{rA}(u,x^A) + O(r^{-4}),\qquad &T_{AB} &= r^{-1}\gamma_{AB} \hat T (u,x^A) + O(r^{-2}).
\end{align}
\end{subequations}
This is obviously satisfied by compact sources for which $T_\mn=0$ outside some finite radius. Our construction however allows for more general configurations involving electromagnetic radiation. The metric field in Bondi gauge reads \emph{on-shell} as 
\begin{align}
\nonumber ds^2=& - c^2 \,du^2 - 2 c\, dudr+r^2 \gamma_{AB}\,dx^Adx^B \\
\nonumber&+\dfrac{2G m}{c^2 r} du^2+r\, C_{AB}\,dx^A dx^B+D^B C_{AB} \,c dudx^A \nonumber
\end{align}
\begin{align}
&+\left( \dfrac{c}{16}\,C_{AB}C^{AB}+\frac{2 \pi G}{c} \hat T_{rr}\right)\frac{1}{r^2}dudr+\dfrac{1}{r}\left[\dfrac{4G}{3c^2}\bar N_A-\dfrac{c}{8}\pd_A\left(C_{BC}C^{BC}\right)\right]dudx^A\nonumber \\
&+\left( \dfrac{1}{4}\gamma_{AB}\,C_{CD}C^{CD}+\mathcal D_{AB} \right) \,dx^Adx^B +\text{(subleading terms in $r$)}.\label{metricBondi}
\end{align}
The stress-energy conservation imposes
\bea
\p_u \hat T_{rA}=c D_A \hat T,\qquad \p_u \hat T_{rr} = -2 c \hat T,
\eea
and Einstein's equations further imply $\hat T_{rA} = -\frac{1}{2}D_A \hat T_{rr}-\frac{1}{8\pi}D^B \mathcal D_{AB}$. Here, $\gamma_{AB}$ is the unit metric on the sphere and all indices are raised and lowered using this metric. The symmetric tracefree field, $C_{AB}(u,x^A)$, is the \emph{Bondi shear} and it contains the transverse and traceless gravitational radiation. Its retarded time derivative, $\dot{C}_{AB}$, is the \emph{Bondi news}; $\mathcal D_{AB}$ is a conserved traceless tensor $\gamma^{AB}\mathcal D_{AB}=\p_u \mathcal D_{AB} = 0$; $m(u,x^A)$ is the \emph{Bondi mass aspect} and $\bar N_A(u, x^A)$ is \emph{Bondi angular momentum aspect} as defined in \cite{Pasterski:2015tva}, which is a convenient definition for writing down the metric \eqref{metricBondi}. We use the following conventions: $m$ has dimension of energy, $\bar N_A$ has dimension of angular momentum, $C_{AB}$ has dimension of length and $\hat T$ has dimension of mass. Einstein's equations then reduce to the following three constraint equations: 
\begin{subequations}\label{constr}\begin{align}
\p_u m &= -4 \pi  c \hat T_{uu} - \frac{c^3}{8 G}\dot{C}_{AB}\dot{C}^{AB}+\frac{c^4}{4G}D_A D_B \dot{C}^{AB},\label{constr1} \\
\p_u \bar N_A &= -8 \pi c \left(\hat T_{uA}+\frac{c}{4}D_A \hat T \right) + \p_A m + \frac{c^4}{4G}D^B (D_A D^C C_{BC}-D_B D^C C_{AC})\nonumber \\
& \quad +\frac{c^3}{4G}D_B \left(\dot{C}^{BC} C_{CA}\right)+\frac{c^3}{2G}D_B \dot{C}^{BC} C_{CA}. \label{constr2}
\end{align}
\end{subequations}

\subsection{Supertranslations and super-Lorentz transformations}
\label{sec:Bondi}

Bondi gauge is preserved by exactly two families of smooth diffeomorphisms: supertranslations and Diff$(S^2)$ super-Lorentz transformations \cite{Campiglia:2014yka,Campiglia:2015yka}. We will refer to the total group of residual diffeomorphisms in Bondi gauge as the extended BMS group. The exact form of the infinitesimal generators can be found, \emph{e.g.}, in \cite{Barnich:2010eb,Compere:2018ylh}, They schematically read as 
\begin{subequations}\label{residual diff}
\begin{align}
\xi_T&=T(x^A)\pd_u+\cdots  &\text{(supertranslations)}\label{supertranslations};\\
\xi_Y &=Y^A(x^B)\p_A + \cdots &\text{(super-Lorentz transformations)}\label{superLorentz};
\end{align}
\end{subequations}
where  $T$ and $Y^A$ respectively denote an arbitrary function and vector on the sphere. The dots in Eqs.~\eqref{residual diff} refer to additional terms required to preserve Bondi gauge. What is important for our purposes is that the symmetries are completely determined by their arguments $T, Y_A$. The action of these diffeomorphisms on the metric can be found, \emph{e.g.}, in \cite{Barnich:2010eb,Hawking:2016sgy,Compere:2018ylh,Compere:2019qed}. 

The four supertranslations, whose generator $T$ obeys
\bea
D_A D_B T + \frac{1}{2}\gamma_{AB} \Delta T = 0,\label{propT}
\eea
\emph{uniquely} define the Poincar\'e translations. Here $\Delta =D_C D^C$ is the Laplacian on the unit 2-sphere. The generator $T$ is given by a linear combination of the $\ell=0,1$ spherical harmonics. Time translation is associated with $T = 1$. Spatial translations are associated with $T=-\frac{1}{c}\, n_i$ where $n_i$, with $i=1,2,3$, are the three components of the unit vector $\vec{n}=(\sin\theta \cos\phi,\sin\theta \sin\phi,\cos\theta)$.\footnote{Indeed, in retarded coordinates, $\partial_{i} = -\frac{1}{c}n_i\partial_{u} + n_i \partial_{r} + \frac{1}{r}P_{il} \frac{\partial}{\partial n_l}$ where $P_{ij}=\delta_{ij}-n_i n_j$.} We will call \emph{proper} supertranslations the supertranslations that are not the time and spatial translations.

The six Diff$(S^2)$ super-Lorentz transformations that preserve the boundary metric of the sphere, or equivalently, whose generator $Y^A$ obeys the conformal Killing equations on the unit sphere
\bea
D_A Y_B + D_B Y_A = \gamma_{AB}D^C Y_C,\label{confK}
\eea
\emph{uniquely} define the six Lorentz asymptotic symmetries projected into the celestial sphere that form a $so(1,3)$ subgroup of diff$(S^2).$\footnote{Also note that $(D_A D_B + \gamma_{AB}) D_C Y^C = 0= (\Delta + 1)Y_A$ and $[D^B,D_A]Y_B = Y_A$.} Within the solutions to \eqref{confK}, the three rotations are uniquely defined as the divergence-free subgroup $D_A Y^A = 0$, while the 3 boosts are uniquely defined as the curl-free generators $\eps^{AB}\p_A Y_B = 0$. The generator $Y^A$ is given by a linear combination of the $\ell=1$ spherical harmonics. We will call \emph{proper} super-Lorentz transformations the super-Lorentz transformations that are not the Lorentz transformations.

A general super-Lorentz generator is parametrized by an arbitrary vector on the sphere, which can be uniquely decomposed as 
\bea
Y^A = \eps^{AB}\pd_B\Phi+\gamma^{AB}\pd_B\Psi. \label{decY}
\eea
In Minkowski spacetime, the Lorentz boosts are given by $\xi_{(i)}=\frac{1}{c}x_i\pd_t+ct\pd_i$. When written in retarded coordinates, one can read off the leading components as $r \rightarrow \infty$ on the sphere as $Y^A=\gamma^{AB}\pd_B n_i$. Similarly, a rotation is given in Minkowski spacetime by $\xi_{(i)}=\eps_{ijk}x_j\pd_k$ and its leading angular components in retarded coordinates are $\eps^{AB}\pd_B n_i$. We infer that the asymptotic rotations and boosts are parametrized by $\Phi \sim n_i$ and $\Psi \sim n_i$, respectively. Therefore, the decomposition \eqref{decY} provides a generalization of those notions and hence we call them super-rotations and super-boosts, respectively. Explicitly, we decompose the super-Lorentz residual transformations \eqref{residual diff} as
\begin{subequations}\label{superLorentz diff}
\begin{align}
\xi_\Phi&=\eps^{AB}\pd_B\Phi \,\pd_A+\cdots &\text{(superrotations)}\label{superrotations};\\
\xi_\Psi&=\gamma^{AB}\pd_B\Psi \,\pd_A+\cdots &\text{(superboosts)}\label{superboosts}.
\end{align}
\end{subequations}

Together, the translations, the Lorentz asymptotic symmetries and the proper supertranslations form the BMS group which is the asymptotic symmetry group of asymptotically flat spacetimes of the form \eqref{metricBondi}. Such asymptotic symmetries obey two fundamental properties: they are associated with finite surface charges, and their action as diffeomorphisms preserves the set of metrics. The proper super-Lorentz transformations are \emph{not} asymptotic symmetries of the set of metrics \eqref{metricBondi}, because they do not preserve the fall-off conditions of the Bondi metric  \eqref{metricBondi} upon action as diffeomorphisms.\footnote{For the construction of an extended phase space of metric admitting super-Lorentz transformations as asymptotic symmetries, see \cite{Barnich:2010eb,Barnich:2011mi,Barnich:2016lyg,Compere:2018ylh,Flanagan:2019vbl}.} Yet, they define finite surface charges for the set of metrics \eqref{metricBondi} and, therefore, they are physically relevant for standard asymptotically flat spacetimes \cite{Flanagan:2015pxa,Compere:2016hzt,Hawking:2016sgy}, as we will detail below. They are particular instances of \emph{external} or \emph{outer} symmetries, which is a more general concept than asymptotic symmetries \cite{Compere:2016jwb,Compere:2017wrj}.

\section{The BMS flux-balance laws}
\label{sec:BMS}

In this section, we present our definition of the extended BMS charges and derive their evolution in terms of retarded time. We will always refer to the flux-balance laws of extended BMS charges as the BMS flux-balance laws. 

\subsection{Supermomenta, super-angular momenta and super-center-of-mass}

The surface charges $\mathcal P_T$ associated with the supertranslations $T$ are called the supermomenta. The surface charges $\mathcal J_\Phi$ and $\mathcal K_\Psi$ associated with the superrotations (labelled by $\Phi $) and the superboosts (labelled by $\Psi$) are, respectively, the super-angular momenta and super center-of-mass charges. We call all these charges the \emph{BMS charges}.  They are built from a complete set of potentials in Bondi gauge that admit a retarded time evolution constrained by Einstein's equations. They are defined as
\begin{subequations} \label{BMScharge}
\begin{empheq}[box=\widefbox]{align}
\mathcal P_T &\equiv\frac{1}{c} \oint_S  T\,  m, \label{convPi}\\
\mathcal J_\Phi &\equiv  \frac{1}{2} \oint_S \eps^{AB}\pd_B\Phi N_A,\label{defJ} \\ 
\mathcal K_\Psi &\equiv \frac{1}{2c} \oint_S  \gamma^{AB}\pd_B\Psi N_A\label{defK}
\end{empheq}
\end{subequations}
where $m$ and $N_A$ are the Bondi mass and angular momentum aspects, respectively. We adjusted the factors of $c$ such that the momentum has dimension mass-times-velocity and the center-of-mass charge has dimension mass-times-length. The angular momentum has its canonical dimension. For $T,\Phi$ and $\Psi$, consisting of linear combinations of $\ell=0,1$ harmonics, the BMS charges \eqref{BMScharge} reduce to the ten Poincar\'e charges. More precisely, we associate the energy (divided by $c$) with $T=1$, the momentum with $T=n_i$ (the momentum is associated with $-\p_i$), the angular momentum with $\Phi = -n_i$ (the angular momentum is associated with $-\p_\varphi$) and the center-of-mass with $\Psi=n_i$. For $T,\Phi$ and $\Psi$ consisting of linear combinations of higher $\ell \geq 2$ spherical harmonics, the BMS charges \eqref{BMScharge} are the \emph{proper BMS charges}. 
While all the literature agrees with the definition of Bondi mass aspect, the definitions of super-angular momentum and super-center-of-mass are ambiguous: they require a prescription for defining the Bondi angular momentum aspect. We will scrutinize the different prescriptions in Section \ref{sec:comparison}. In this paper, we prescribe the Bondi angular momentum aspect $N_A$ as 
\begin{align}
N_A &= \bar  N_A - u \p_A m -\frac{c^3}{4G} C_{AB} D_C C^{BC} -\frac{c^3}{16G} \partial_A \left(C_{BC} C^{BC}\right) \nonumber\\
& \quad + \frac{uc^4}{4G} D_B D^B D^C C_{AC}- \frac{uc^4}{4G} D_B D_A D_C C^{BC} \label{NA}
\end{align}
where $\bar N_A$ is defined from the metric expansion \eqref{metricBondi}. This prescription was found in \cite{Compere:2018ylh}\footnote{It was denoted as $\hat N_A$ and defined in Eq.~(5.52).} by requiring that the Ward identity associated with super-Lorentz symmetries reproduces the standard form of the subleading soft graviton theorem \cite{Cachazo:2014fwa}. In Section \ref{sec:flux} of this paper, we will further prove that the definition \eqref{NA} exactly leads to the flux-balance law for the angular momentum as computed in \cite{Thorne:1980ru} \emph{without} any additional divergences or total time derivatives. Moreover, it will lead to the flux-balance law for the center-of-mass that matches with the post-Newtonian derivation of \cite{Kozameh:2017qiw}. Furthermore, the center-of-mass flux will be related to the symplectic flux, which parallels the prescription used to defining the angular momentum flux \cite{Ashtekar:1981bq}.

The term $-u \p_A m$ in Eq.~\eqref{NA} cancels the linear $u$ divergence present in $\bar N_A$ for non-radiative configurations, as shown in \cite{Hawking:2016sgy}, and as we will rederive in Section~\ref{sec:init}. The super-center-of-mass charge $\mathcal K_\Psi$ is therefore exactly conserved for non-radiative configurations, even in the presence of supermomentum. In particular, the center-of-mass charge $\mathcal K_i$, defined for $\Phi = n_i$,  is conserved in the presence of momentum and therefore physically measures mass times the initial (\emph{i.e.}, at $u=0$) center-of-mass position. The center-of-mass that evolves linearly as a function of momentum in the absence of radiation is instead given by $\mathcal G_i \equiv \mathcal K_i + u \mathcal P_i$. We can generalize this definition to all superboosts by defining the \emph{comoving} super-center-of-mass, 
\begin{empheq}[box=\widefbox]{align}
\mathcal G_\Psi \equiv \frac{1}{2c} \oint_S  \gamma^{AB}\pd_B\Psi \left( N_A + u \p_A m \right).
\end{empheq}
The comoving super-center-of-mass $\mathcal G_\Psi $ is related to the super-center-of-mass by 
\bea
\mathcal G_\Psi = \mathcal K_\Psi + u\, \mathcal P_{T = -\frac{1}{2} \Delta \Psi} \label{defG}.
\eea
For boosts, $\Psi = n _i$, $-\frac{1}{2}\Delta n_i = n_i$ and we recover the standard relationship $\mathcal G_i \equiv \mathcal K_i + u \mathcal P_i$. Finally note that the shift $N_A \mapsto N_A - u \p_A m$ does not affect the super-angular momentum because $\eps^{AB}\p_A \p_B \Phi = 0$.

\subsection{The BMS flux-balance laws}

The BMS flux-balance laws are simply obtained by taking the $u$-derivative of the BMS charges \eqref{BMScharge} and using the constraint equations \eqref{constr} derived from the Einstein's equations.

The evolution of the Bondi mass aspect $m$ is given by Eq.~\eqref{constr1}. One can multiply each side of this equation by an arbitrary function $T(x^A)$ over the sphere. Upon integration and using the definition of the supermomentum \eqref{convPi}, one obtains the supermomentum flux-balance law 
\begin{empheq}[box=\widefbox]{align}
\dot {\mathcal P}_T - \frac{c^3}{4 G}   \oint_S  T ~D_A D_B \dot C^{AB}   &=   - \frac{c^2}{8G}\oint_S T\;  \left( \dot C_{AB} \dot C^{AB} + \frac{32\pi G}{c^2}\hat T_{uu}\right).\label{fluxPT}
\end{empheq}
This flux-balance law has been well-studied and derived in many references, including \cite{1962RSPSA.269...21B,1962RSPSA.270..103S,Payne:1982aa,Strominger:2014pwa}. The term linear in the shear tensor $ C_{AB}$ gives the soft contribution to the Weinberg's leading soft graviton theorem upon quantization. The quadratic term in the shear as well as the matter contribution are called the \emph{hard} terms, as they contribute to the energy flux.

The evolution of the Bondi angular momentum aspect $\bar N_A$ is given by Eq.~\eqref{constr2}. The definitions of the super-angular momentum and super-center-of-mass charges \eqref{defJ}-\eqref{defK} involve $N_A$ given in Eq.~\eqref{NA}. It is a simple matter of algebra to write down the retarded time evolution of $\mathcal J_\Phi$ and $\mathcal K_\Psi$ or, equivalently, $\mathcal G_\Psi$.  The answer is most easily expressed as follows. We first introduce the bilinear \emph{hard super-Lorentz} operator \cite{Campiglia:2016hvg}
\begin{align}\label{curly H def}
\mathcal{H}_A (\dot C,C) &\equiv \frac{1}{2} \partial_A \left(\dot C_{BC} C^{BC}\right) - \dot C^{BC} D_A C_{BC} + D_B \left(\dot C^{BC} C_{AC} - C^{BC} \dot C_{AC}\right)\,,
\end{align}
and the linear \emph{soft superrotation} operator 
\bea \label{defSA}
\mathcal{S}_A (\dot C) \equiv  \Delta  D^C \dot C_{AC}-D_B D_A D_C \dot C^{BC}.\label{softSA}
\eea
The constraint \eqref{constr2} can then be rewritten in terms of $N_A$ as
\begin{align}
\partial_u N_A + u \p_A \dot m &=\frac{c^3}{4G}\mathcal H_A(\dot C,C)+\frac{uc^4}{4G}{ \mathcal S}_A(\dot C) -8\pi c \, \bar T_{uA}
\label{duNa}
\end{align}
where $\bar T_{uA} \equiv \hat T_{uA} + \frac{c}{4} D_A \hat T - \frac{u}{2}\hat T_{uu}$. Eq.~\eqref{duNa} is in fact a rewriting of Eq.~(5.53) of \cite{Compere:2018ylh}, complemented with the matter contributions. After contraction with $Y^A$ and integration over the sphere, one obtains the flux-balance laws for the super-angular momentum and super-center-of-mass.

Let us now obtain the simplest possible form for these flux-balance laws. Here, we first observe that the soft superrotation operator $\mathcal{S}_A$ in Eq.~\eqref{defSA} does not contribute to the flux-balance law for the super-center-of-mass. Indeed, contracting Eq.~\eqref{defSA}, respectively, with a superrotation $Y^A_{\Phi} = \eps^{AB}\pd_B\Phi$ and a superboost $Y^A_{\Psi} = \gamma^{AB}\pd_B\Psi$, one can show after integration by parts that 
\begin{subequations}
\begin{align}
\oint_S \eps^{AB}\pd_B \Phi ~{\mathcal S}_A(\dot C)  &= \oint_S  \Delta \Phi  D_A D_B \left(\eps^A_{\;\; C} \dot C^{CB}\right),\label{softJ}\\
\oint_S \gamma^{AB}\pd_B \Psi ~{\mathcal S}_A(\dot C)  &= 0.\label{softK} 
\end{align}
\end{subequations}
The operator $\mathcal S_A$ therefore deserves its name as the soft superrotation operator. We also observe that the soft contribution to the super-angular momentum flux-balance law \eqref{softJ} is similar to the soft contribution to the supermomentum flux-balance law \eqref{fluxPT}. The difference amounts to a parity flip of the Bondi news: the soft supertranslation term depends upon the parity-even part, while the soft superrotation term depends upon the parity-odd part of the news. The structure can be made explicit by performing the decomposition of the Bondi shear into its two polarization modes\footnote{Note that under a supertranslation, $C^+ \mapsto C^+ + T$, while $C^-$ is invariant.}
\bea
C_{AB} = -2 D_A D_B C^++\gamma_{AB}\Delta C^+ +2 \eps_{C(A}D_{B)}D^C C^-.\label{Cpm}
\eea
The BMS flux-balance laws then take the simpler form 
\begin{subequations} 
\begin{align}
\dot {\mathcal P}_T + \frac{c^3}{4 G} \oint_S  T (\Delta + 2) \Delta \dot C^+  &=  - \frac{c^2}{8G}\oint_S T\;  \left( \dot C_{AB} \dot C^{AB} + \frac{32\pi G}{c^2}\hat T_{uu}\right) ,\label{fluxPT3}\\
\dot {\mathcal J}_\Phi +u \frac{c^4}{4G}  \oint_S \left(-\frac{1}{2}\Delta\Phi \right)  (\Delta + 2) \Delta \dot C^-  &= + \frac{c^3}{8G}\oint_S  \eps^{AB}\pd_B\Phi \; \left( \mathcal H_A(\dot C,C)-\frac{32\pi G}{c^2} \bar T_{uA} \right) ,  \label{fluxJphi} \\
\dot {\mathcal K}_\Psi + u\, \dot{\mathcal P}_{T= -\frac{1}{2}\Delta \Psi} &=+ \frac{c^2}{8G}\oint_S \gamma^{AB}\pd_B\Psi  \left( \mathcal H_A(\dot C,C)-\frac{32\pi G}{c^2} \bar T_{uA} \right).\label{sboostlaw2}
\end{align}
\end{subequations}
Notice that using Eq.~\eqref{defG}, we can rewrite 
\bea\label{GK}
\dot {\mathcal K}_\Psi + u\, \dot{\mathcal P}_{T= -\frac{1}{2}\Delta \Psi} = \dot{\mathcal G}_\Psi - {\mathcal P}_{T= -\frac{1}{2}\Delta \Psi} .
\eea

Thanks to this simplified form, the physical content of the soft contributions to the BMS flux-balance laws is now apparent. The supermomentum flux-balance law is associated with the displacement memory effect which is caused by a permanent displacement of $C^+$ between non-radiative regions due to null radiation reaching null infinity \cite{Zeldovich:1974aa,Payne:1982aa,0264-9381-9-6-018,Blanchet:1987wq,Christodoulou:1991cr,Blanchet:1992br,Bieri:2013ada}. The super-angular momentum flux-balance law is associated with the super-angular momentum memory effect dubbed the ``spin memory effect'' \cite{Pasterski:2015tva,Nichols:2017rqr,Flanagan:2018yzh,Himwich:2019qmj}. It is caused by an accumulation $\int_{u_1}^{u_2}u \p_u C^-$ between the initial and final retarded time of the operator $u \p_u$ acting on the parity-odd radiative polarization mode. Though the spin memory effect is not clearly a memory effect (\emph{i.e.}, an effect depending only on the initial and final states) using variables in Bondi gauge, it is clearly a memory effect once rewritten in canonical/harmonic gauge \cite{Himwich:2019qmj}. The only soft contribution to the super-center-of-mass flux-balance law \eqref{sboostlaw2} arises from the soft contribution to the supermomentum, with $T= - \frac{1}{2}\Delta\Psi$, and is proportional to $u \p_u C^+$. The ``super-center-of-mass memory effect'', or ``center-of-mass memory effect'' in short, arises from an accumulation $\int_{u_1}^{u_2}u \p_u C^+$ between the initial and final retarded time of the operator $u \p_u$ acting on the parity-even radiative polarization mode \cite{Nichols:2018qac}. We will come back to the memory effects in Section \ref{sec:glob}, where we write down the time-integrated form of the flux-balance laws.

Let us now simplify the right-hand side of the super-angular momentum and super-center-of-mass flux-balance laws. Using integration by parts and Eq.~\eqref{confK}, we observe that the first term in the hard super-Lorentz operator \eqref{curly H def} does not contribute to the super-angular momentum flux-balance law (though we will keep it to simplify the integrand), while the third term does not contribute to super-center-of-mass flux-balance law. In addition, not all quadratic terms are independent from each other. Indeed, for any pair of symmetric tracefree bidimensional tensors $X_{AB}$, $Y_{AB}$, we have
\bea
X_{AC} D_B Y^{BC} = X^{BC} D_A Y_{BC} - X^{BC} D_B Y_{AC}.
\eea
We can therefore choose two quadratic operators as a basis and express the hard contributions in terms of them. It is convenient to define 
\begin{subequations}\label{T1T2}
\begin{align}
T^{(1)}_A(\dot C,C) \equiv \frac{1}{2} ( \dot C^{BC}D_A  C_{BC} - C^{BC}D_A \dot C_{BC}) ,\\
T^{(2)}_A(\dot C, C) \equiv \frac{1}{2} ( \dot C^{BC}D_B C_{AC} -  C^{BC}D_B \dot C_{AC}) .
\end{align}
\end{subequations}
Following the steps mentioned above, the BMS flux-balance laws finally read as
\begin{subequations}\label{BMSfluxes}
\begin{empheq}[box=\fbox]{align}
\dot {\mathcal P}_T + \frac{c^3}{4 G} \oint_S  T~ (\Delta + 2) \Delta \dot C^+  &= - \frac{c^2}{8G}\oint_S T\;  \left( \dot C_{AB} \dot C^{AB} + \frac{32\pi G}{c^2}\hat T_{uu}\right)  ,\label{fluxPT2}\\ 
\dot {\mathcal J}_\Phi -u \frac{c^4}{8G}  \oint_S \Delta\Phi   (\Delta + 2) \Delta \dot C^-  &=  + \frac{c^3}{8G}\oint_S \eps^{AB}\pd_B\Phi \left( - 3 T^{(1)}_A(\dot C,C)  + 4 T^{(2)}_A(\dot C,C)  \right) \nonumber\\
&\quad -4 \pi c\oint_S  \eps^{AB}\pd_B\Phi ~ \bar T_{uA}, \label{fluxJ2}\\
\dot {\mathcal K}_\Psi + u\, \dot{\mathcal P}_{T= -\frac{1}{2}\Delta \Psi}  &=  - \frac{c^2}{8G} \oint_S \gamma^{AB}\pd_B \Psi ~T^{(1)}_A( \dot C, C)\nonumber\\
&\quad -4 \pi\oint_S \gamma^{AB}\pd_B\Psi~  \bar T_{uA}.\label{sboostlaw}
\end{empheq}
\end{subequations}
While these laws have been written down for various definitions of BMS charges, it is the first time that they are written for the definition of the super-Lorentz charges \eqref{defJ}-\eqref{defK} in simplified form. A comparison with the literature will be provided in Section \ref{sec:comparison}. The proper supermomentum, super-angular momentum and super-center-of-mass flux-balance laws will be expanded in post-Newtonian form in Sections \ref{sub:supermomenta}, \ref{sec:superrot} and \ref{sec:superboost}. Finally note that one could also absorb the soft terms into the supermomenta and thereby defining \emph{dressed} BMS supermomenta as done for electric asymptotic charges in \cite{Mirbabayi:2016axw}. 

\subsection{Poincar\'e flux-balance laws}  \label{Poincare fluxes}

In the special case of Poincar\'e generators, the BMS flux-balance laws simplify. All the soft terms linear in $C_{AB}$ vanish for Poincar\'e generators which correspond to the $\ell = 0,1$ harmonics of the functions $T,\Phi,\Psi$. This is because they are zero modes of the operator $\Delta (\Delta + 2)$ in Eq.~\eqref{BMSfluxes}.

According to our convention in \eqref{convPi}, the energy (divided by $c$) is canonically associated to $T=1$ and the linear-momentum is canonically associated to $T= n_i$.
The flux-balance laws of energy and momentum read, respectively, as 
\begin{subequations}\label{EPflux}
\begin{align}
\dot {\mathcal E} &\equiv  c \, \dot{\mathcal P}_{T=1}=  - \frac{c^3}{8G} \oint_S \left( \dot C_{AB} \dot C^{AB} + \frac{32\pi G}{c^2}\hat T_{uu}\right) , \label{fluxE}\\
\dot {\mathcal P}_i &\equiv  \dot{\mathcal P}_{T=n_i}= - \frac{c^2}{8G} \oint_S \left( \dot C_{AB} \dot C^{AB} + \frac{32\pi G}{c^2}\hat T_{uu}\right) n_i.
\label{fluxP}
\end{align}
\end{subequations}

According to Eqs.~\eqref{defJ} and \eqref{defK}, the angular momentum and center-of-mass are canonically associated to $\Phi = -n_i$ and  $\Psi = n_i$ respectively. Their flux-balance laws explicitly read
\begin{subequations}\label{JKflux}
\begin{align}
\dot {\mathcal J}_i&\equiv \dot{\mathcal J}_{\Phi=-n_i} =  -\frac{c^3}{8G}  \oint_S  \eps^{AB}\pd_B n_i \left( - 3 T^{(1)}_A(\dot C,C)  + 4T^{(2)}_A(\dot C,C) - \frac{32 \pi G}{c^2} \bar T_{uA} \right),\label{fluxJ} \\
\dot {\mathcal K}_i  +u \, \dot {\mathcal P}_i &= \dot {\mathcal G}_i  - {\mathcal P}_i = - \frac{c^2}{4G} \oint_S \gamma^{AB}\pd_B n_i \left(  \frac{1}{2} T^{(1)}_A( \dot C,C)+ \frac{16 \pi G}{c^2} \bar T_{uA}\right)  \label{fluxK}
\end{align}
\end{subequations}
where ${\mathcal K}_i \equiv {\mathcal K}_{\Psi=n_i}$, ${\mathcal G}_i \equiv {\mathcal G}_{\Psi=n_i}$. 
These ten Poincar\'e flux-balance laws will be expanded in terms of symmetric tracefree radiative multipoles in Section \ref{PN Poincare}.

\subsection{On the definitions of angular momentum and center-of-mass}
\label{sec:comparison}

Several definitions of angular momentum and center-of-mass of asymptotically flat spacetimes have been proposed. Here, we summarize some of these definitions, relate them to each other and discuss their properties. Let us define
\begin{subequations}\label{genJ}
\begin{align}
\mathcal J^{(\alpha )}_i &\equiv   -\frac{1}{2}\oint_S \eps^{AD}\p_D n_i \,\left(\bar N_A -\frac{\alpha c^3}{4G} C_{AB} D_C C^{BC}\right) ,\\ 
\mathcal K^{(\alpha,\beta )}_i &\equiv  + \frac{1}{2}\oint_S \gamma^{AD}\p_D n_i \,\left(\bar N_A - u \p_A m -\frac{\alpha c^3}{4G} C_{AB} D_C C^{BC} - \frac{\beta c^3}{16 G}\p_A \left(C_{BC}C^{BC}\right)\right) ,\\ 
\mathcal G^{(\alpha ,\beta)}_i &\equiv +  \frac{1}{2}\oint_S \gamma^{AD}\p_D n_i \,\left(\bar N_A -\frac{\alpha c^3}{4G} C_{AB} D_C C^{BC} - \frac{\beta c^3}{16 G}\p_A \left(C_{BC}C^{BC}\right)\right), 
\end{align}
\end{subequations}
with $\alpha$, $\beta$ arbitrary. Using Eq.~\eqref{NA}, our definition corresponds to $\alpha=\beta=1$, which matches with the one of Barnich-Troessaert \cite{Barnich:2011mi} and subsequently \cite{Flanagan:2015pxa,Distler:2018rwu,Compere:2018ylh}. Instead, the angular momentum and center-of-mass as defined by Pasterski-Strominger-Zhiboedov \cite{Pasterski:2015tva} and subsequently \cite{Hawking:2016sgy} corresponds to $\alpha =\beta= 0$. We can also compare our expression for the angular momentum flux \eqref{fluxJ} with equation Eq.~(4.11) of \cite{Bonga:2018gzr} derived from the Landau-Lifshitz pseudo-tensor. Their expression for the flux of angular momentum  is of the form \eqref{fluxJ2}, but with $T^{(1)}_A(\dot C,C)=\dot C^{BC}D_A C_{BC}$ and $T^{(2)}_A(\dot C,C)=\dot C^{BC}D_B C_{AC}$ instead of Eq.~\eqref{T1T2}. Their angular momentum flux coincides with $\mathcal J^{(\alpha=3)}$ in Eq.~\eqref{genJ}.

For any $\alpha$, one infers from Eq.~\eqref{fluxJ2} that the flux of angular momentum $\mathcal J^{(\alpha )}_i$ can be written purely in terms of the radiative data, \emph{i.e.}, the shear and the news, and has no explicit dependence on the Coulombic data, which agrees with \cite{Bonga:2018gzr}.

In order to relate the angular momentum to the one defined by Dray and Streubel \cite{Dray:1984rfa}, we need to recall the definition of the symplectic structure \cite{Crnkovic,Ashtekar:1981bq,Lee:1990nz}. It is defined as $\omega(\delta_1 g,\delta_2 g) \equiv \delta_1 \Theta(\delta_2 g) - \delta_2 \Theta(\delta_1 g)$ where $\Theta(\delta g)$ is the boundary term obtained after varying the Einstein-Hilbert Lagrangian. Expanding this symplectic structure close to null infinity, and integrating over the sphere, one obtains the finite symplectic form at null infinity
\bea\label{symplectic}
\omega_{\mathcal I^+}(\delta_1 g,\delta_2 g) = \frac{c^3}{8G}\oint_S \left( \delta_1 \dot C^{AB} \delta_2 C_{AB}-\delta_2 \dot C^{AB} \delta_1 C_{AB}\right) . 
\eea
The action of the Lie derivative with respect to a supertranslation or Lorentz transformation on the metric $\mathcal L_{Y,T} g$ gives the induced action on the shear $C_{AB}$ and news $\dot C_{AB}$ given by 
\begin{subequations}
\begin{align}
\delta_{T,Y} C_{AB} &= \left(T+\frac{u}{2}D_A Y^A\right)\dot C_{AB}+\mathcal L_Y C_{AB}-\frac{1}{2} D_C Y^C C_{AB}-2 D_A D_B T + \gamma_{AB}\Delta T,\\
\delta_{T,Y} \dot C_{AB} &= \left(T+\frac{u}{2}D_A Y^A\right) \ddot C_{AB}+\mathcal L_Y \dot C_{AB}. \label{deltaNAB}
\end{align}
\end{subequations}
After some algebra, one can then rewrite the flux of angular momentum \eqref{fluxJ} associated with the vector $Y^A=-\eps^{AB}\p_B n_i$ without matter flux ($\bar T_{u A} = 0$) as
\begin{align}
\dot {\mathcal J}_i  &= - \frac{c^3}{16G} \oint_S \left(\dot C^{AB} \mathcal L_Y C_{AB} - \mathcal L_Y \dot C^{AB} C_{AB}\right) \nonumber\\
&= -\frac{1}{2}\omega_{\mathcal I^+}(g,\mathcal L_Y g) \label{JsF}.
\end{align}
In the last step, we have used the fact that the following term vanishes for divergence-free vectors including rotations,
\bea
\oint_S C^{AB} \mathcal L_Y C_{AB}  = 0. 
\eea 
The angular momentum flux \eqref{JsF} is precisely the one prescribed by Ashtekar and Streubel \cite{Ashtekar:1981bq}. The corresponding angular momentum surface charge \cite{Dray:1984rfa,Shaw:1984sfa} thus corresponds to $\mathcal J_i^{(\alpha=1)}$ in Eq.~\eqref{genJ}. 

Covariant phase space or Hamiltonian charges \cite{Wald:1993nt,Iyer:1994ys,Barnich:2001jy} are well-known to be non-integrable for radiating spacetimes. This leads to the necessity of a prescription to define the canonical charges. Such a prescription was proposed by Wald and Zoupas \cite{Wald:1999wa} that uniquely leads to the Ashtekar-Dray-Streubel angular momentum \cite{Ashtekar:1981bq,Dray:1984rfa}. In Section \ref{sec:flux}, we will further show that the definition of angular momentum for $\alpha = 1$ leads to the angular momentum flux-balance law written in Thorne \cite{Thorne:1980ru}. The angular momentum prescriptions in this paper and  \cite{Thorne:1980ru,Ashtekar:1981bq,Dray:1984rfa,Wald:1999wa,Barnich:2010eb,Barnich:2011mi,Distler:2018rwu,Compere:2018ylh} are therefore all equivalent to each other ($\alpha =1$), while the angular momentum (and corresponding angular momentum flux) prescribed in \cite{Pasterski:2015tva,Hawking:2016sgy} ($\alpha =0$) or \cite{Bonga:2018gzr} ($\alpha =3$) are distinct. 

Let us conclude with a remark on the uniqueness. The motivation of \cite{Ashtekar:1981bq,Dray:1984rfa,Wald:1999wa} was to obtain a definition of angular momentum as an integral of covariant fields over the celestial sphere whose flux vanishes for non-radiative configurations. Now, it is clear that the definitions \eqref{genJ} are all locally defined in terms of the metric fields in Bondi gauge for any $\alpha$, $\beta$. In addition, for non-radiative configurations, the fluxes of $\mathcal J^{(\alpha)}$ and $\mathcal K^{(\alpha,\beta)}$ are vanishing for any $\alpha$, $\beta$, as will be clear from Section \ref{sec:glob}. There is therefore (at least) a one-parameter ambiguity in the definition of the angular momentum and a two-parameter ambiguity in the definition of center-of-mass when one only imposes that these charges are built locally from the Bondi fields and that their fluxes vanish for non-radiative configurations. Note that the subleading soft graviton theorem \cite{Cachazo:2014fwa} is obtained from quantizing the super-Lorentz flux-balance laws, independently on how we shift the left and right-hand sides of the flux-balance law. The existence of the subleading soft theorem, therefore, does not fix the prescription either. 

What is therefore the most convenient prescription? A natural requirement is that the angular momentum should transform in the standard fashion under translations in non-radiative regions. Translations do not affect the shear, $\delta_{T=n_j} C_{AB}= 0$, and therefore the transformation law of $\mathcal J_i^{(\alpha)}$ is \emph{independent} of $\alpha$. For non-radiative configurations the Bondi angular momentum aspect $\bar N_A$ defined from Eq.~\eqref{metricBondi} changes as
\bea
\bar N_A \mapsto \bar N_A +3 \frac{m}{ c}  D_A T + \frac{T}{c} \p_A m ,\label{tran}
\eea
as can be deduced from, \emph{e.g.}, Eq. (2.24) of \cite{Compere:2018ylh}. The transformation of the angular momentum under a translation in non-radiative regions is then given by
\bea
\delta_{T=n_j} \mathcal J_i^{(\alpha)}= \mathcal P_{T''}, \qquad T'' = -Y^A \p_A T = \eps_{jik}n_k,
\eea
where $Y^A = -\eps^{AB}\p_B n_i$. This leads to the standard Poincar\'e commutator $[\mathcal P_j,\mathcal J_i]=\eps_{jik}\mathcal P_k$, independently of the prescription for $\alpha$. One other natural requirement is to impose the simplest transformation property under supertranslations. For non-radiative configurations, the transformation law of $\bar N_A$ under supertranslations does not admit linear terms in the shear, while all other prescriptions do since the shear transforms under supertranslations. This leads to the preferred choice $\alpha=\beta=0$ which is used in \cite{Pasterski:2015tva,Hawking:2016sgy}. Alternatively, one could impose a natural requirement that the flux of angular momentum does not involve mixed parity terms; see Eq.~\eqref{fluxJalpha} below. This fixes instead $\alpha = 1$ as in \cite{Thorne:1980ru,Ashtekar:1981bq,Dray:1984rfa,Wald:1999wa,Barnich:2010eb,Barnich:2011mi,Distler:2018rwu,Compere:2018ylh}. 
In conclusion, we do not see any convincing argument to prefer either prescription.

It is also appealing to define an intrinsic angular momentum which is independent from supertranslations. Such an intrinsic angular momentum was defined in \cite{Javadinazhed:2018mle} using an implicit dressing procedure; see also \cite{Moreschi:2002ii,Gallo:2014jda} for another construction. We will provide the explicit definition of the supertranslation-invariant intrinsic angular momentum in terms of Bondi fields in Section \ref{sec:glob} for non-radiative configurations. The catch is that this definition is non-local in terms of the Bondi fields, as anticipated in \cite{Javadinazhed:2018mle}.

\section{Multipole expansion of the BMS flux-balance laws}
\label{sec:flux}

Solving the binary problem in General Relativity requires a precise accounting of the energy and angular momentum fluxes radiated by the binary. Building upon the work of \cite{1975ApJ...197..717E,1976ApJ...210..764W}, Thorne \cite{Thorne:1980ru}  summarized the flux-balance law for energy, momentum and angular momentum obtained using either pseudo-tensorial methods or expansions in radiative multipole moments after averaging over oscillations and restricting to sources in the rest frame. The purpose of this section is the provide the multipole expansion of the BMS flux-balance laws \eqref{BMSfluxes} in terms of  radiative multipole moments. This derivation allows to obtain the instantaneous form of the flux-balance laws independently of the nature of the sources in the bulk of spacetime. We will explicit both the Poincar\'e flux-balance laws \eqref{EPflux}-\eqref{JKflux} and the proper supermomentum, super-angular momentum and super-center-of-mass flux-balance laws. 
For simplicity, we will drop the contribution of the matter stress-tensor to the flux-balance laws in this section.

\subsection{Multipole expansion of the radiation field}

Multipole expansions can be expressed in two equivalent ways: either in terms of spherical harmonics or in terms of 
symmetric tracefree (STF) tensors. In this work, we will use STF tensors; see Appendix \ref{app:sph} and \cite{Thorne:1980ru} for conversion formulae to spherical harmonics. We use the convention that STF tensors are written in bold font. The capital index $L$ refers to a set of $\ell$ indices, \emph{i.e.},  $\boldsymbol T_{L}= T_{i_1\cdots i_\ell}$. We denote by $N_L$ the  symmetric product of $\ell$ unit vectors $n_i$. We will write the STF projection of a tensor $T_{i_1\cdots i_\ell}$ as $T_{\langle i_1\cdots i_\ell \rangle}$, as in Eq.~(1.8) of \cite{Thorne:1980ru}. Since all flux-balance laws are expressed in terms of finite tensors tangent to the celestial sphere at null infinity, we can re-express all quantities using the Euclidean embedding of the unit sphere.\footnote{Alternatively, we could use the sphere of radius $r$ but since we are working with finite quantities, we find the use of $r \rightarrow \infty$ unnecessary.}  We define the transverse projector $P_{ij}$ and the traceless transverse projector $P_{ijkl}$ as
\begin{equation}
P_{ij}=\delta_{ij}-n_i n_j, \qquad P_{ijkl}=P_{ik}P_{jl}-\frac{1}{2}P_{ij}P_{kl}.
\end{equation}
For any pair of vectors $X_A$ and $Y_{A}$ defined on the unit sphere, we can re-express $Y^A X_A = Y_i P_{ij}X_j$ and $\eps^{AB}X_{A}Y_{B}=n_i\eps_{ijk}X_{j}Y_{k}$ using the Euclidean embedding of the unit 2-sphere. We can also use the tangential derivative on the unit sphere, 
\bea \label{partialderhat}
\hat \partial_{i} = r P_{ik}\p_k =  P_{il} \frac{\partial}{\partial n_l}.
\eea
This allows to write the integrands of the flux-balance laws \eqref{BMSfluxes} in Cartesian notation. 

The BMS symmetry parameters can be expanded as
\begin{align} \label{BMSsymmSTF}
    T(x^A)&=\sum_{\ell=0}^\infty \boldsymbol{T}_L N_L\,,\qquad \Phi(x^A)=\sum_{\ell=1}^\infty \dfrac{1}{\ell}\bS_L N_L\,,\qquad   \Psi(x^A)=\sum_{\ell=1}^\infty \dfrac{1}{\ell} \bK_L N_L\,
\end{align}
where $\boldsymbol{T}_L$, $\bS_L$ and $\bK_L$ are STF tensors. The Cartesian expressions for the superrotations of parameter $S_L$ and superboosts of parameter $K_L$ are respectively
\begin{subequations}\label{superLorentz Cartesian}
\begin{align}
    \left(Y^{S}\right)_i&=-\eps_{ijk} \,N_{jL-1}\,S_{k L-1}\,,&\text{(superrotation);}\\ 
    \left(Y^{K}\right)_i&=N_{L-1}K_{iL-1}-N_{iL}K_L\,.&\text{(superboost).}
\end{align}
\end{subequations}
In particular, for $\ell=1$ and $S_k=\de_{ki}$ we get the usual generator of rotation on the unit sphere around the $i$-th axis. 

The symmetric and traceless Bondi shear $C_{AB}$ can be expressed in terms of the two sets of radiative multipole moments $\bU_L$ (parity-even) and $\bV_L$ (parity-odd) following \cite{Thorne:1980ru,Blanchet:2013haa} as
\bea
C_{ij}&=P_{ijkl} \chi_{kl}=\chi_{ij}-2n_{(i}\chi_{j)}+\dfrac{1}{2}(\delta_{ij}+N_{ij})\chi \label{CijM}
\eea
where $\chi_{ij}$, $\chi_i \equiv n_i \chi_{ij}$ and $\chi \equiv n_i \chi_{i}$ are respectively given by
\begin{subequations}\label{UV}
\begin{align}
\chi_{ij} &= \sum_{\ell = 2}^{+\infty} a_{\ell} \left( N_{L-2} ~\bU_{ij L-2} - \frac{b_\ell}{c} N_{aL-2} ~\eps_{ab (i} \bV_{j) b L-2} \right),\\
\chi_i &= \sum_{\ell = 2}^{+\infty} a_{\ell} \left( N_{L-1}~ \bU_{i L-1} -\frac{b_\ell}{2c} N_{aL-1}~ \eps_{ab i} \bV_{ b L-1} \right)\,,\\
\chi &= \sum_{\ell = 2}^{+\infty} a_{\ell}~  N_{L}~ \bU_{L} ,
\end{align}
\end{subequations}
with the coefficients $a_\ell$ and $b_\ell$ given by
\begin{equation}
a_{\ell} \equiv \frac{4G}{c^{\ell+2} \ell! },\qquad b_{\ell} \equiv \frac{2\ell}{\ell+1}.
\end{equation}

The Bondi shear was expressed in terms of the parity-even and parity-odd polarizations, respectively, $C^+$ and $C^-$, in Eq.~\eqref{Cpm}. These two polarizations are respectively given as a function of $\bU_L$ and $\bV_L$ as 
\begin{subequations}\label{dictCp}
\begin{align}
	C^+ &= -\sum_{\ell = 2}^\infty \frac{a_\ell}{2\ell (\ell-1)} \bU_{L}  N_L + \bar X^{(0)} + \bar X_i^{(0)}n_i ,\label{cp}\\
	C^- &= -\sum_{\ell = 2}^\infty \frac{a_\ell b_\ell}{2c \ell (\ell-1)} \bV_{L}  N_L+ \tilde X^{(0)} +\tilde X_i^{(0)}n_i \label{cm} 
\end{align}
\end{subequations}
where $\bar X^{(0)}$, $\bar X_i^{(0)}$, $\tilde X^{(0)}$, $\tilde X_i^{(0)}$ are arbitrary because they are zero modes of the differential operators \eqref{Cpm} and they do not appear in the metric.

In the post-Minkowskian formalism \cite{Blanchet:1985sp,Blanchet:1987wq,Blanchet:1992br}, the radiative multipole moments can be expressed in terms of the auxiliary canonical multipole moments $ \boldsymbol M_L$, $ \boldsymbol S_L$ and source multipole moments $ \boldsymbol I_L$, $ \boldsymbol J_L$ as 
\begin{subequations}\label{MS}
\begin{align}
	\bU_{L} &=  \stackrel{(\ell )}{\boldsymbol M}_{L} + \mathcal{O}\left( \frac{G}{c^3}\right) =  \stackrel{(\ell )}{\boldsymbol I}_{L} + \mathcal{O}\left( \frac{G}{c^3}\right),\\
	\bV_{L} &=\stackrel{(\ell )}{\boldsymbol S}_{L} + \mathcal{O}\left( \frac{G}{c^3}\right)= \stackrel{(\ell )}{\boldsymbol J}_{L} + \mathcal{O}\left( \frac{G}{c^3}\right)
\end{align}
\end{subequations}
where the superscript $(\ell)$ denotes $\ell$ derivatives with respect to $u$. Explicit formulae beyond the leading term can be found in Eqs.~(95)--(98) of \cite{Blanchet:2013haa}.

\subsection{Poincar\'e flux-balance laws} \label{PN Poincare}

We will now derive the multipole expansion of the Poincar\'e flux-balance laws derived in the previous Section \ref{Poincare fluxes}.

\paragraph{Energy-momentum flux-balance laws.}
Let us start by deriving the multipolar expansion of the flux of energy and linear-momentum \eqref{EPflux}. 
The computation consists in substituting the quadratic expression $\dot C^{AB}\dot C_{AB}=\dot C^{ij}\dot C_{ij}$, provided in Eq.~\eqref{CC}, and using the explicit integrals in Eq.~\eqref{EPintegrals}. After some algebra, we obtain the \emph{exact} multipolar expansion of the energy-momentum flux-balance laws in terms of the radiative multipoles  
\begin{subequations}\label{FluxesEP}\begin{align}
\dot {\mathcal E} &= - \sum^{+\infty}_{\ell=2} \frac{G}{c^{2\ell+1}}\mu_{\ell} 
\biggl\{\dot {\bU}_{\!L} \dot{\bU}_{\!L} + \frac{b_\ell b_\ell}{c^2}  \dot{\bV}_{\! L}
\dot{\bV}_{\! L}\biggr\} \,,\label{FluxE}\\ 
\dot {\mathcal P_i} &= - \sum^{+\infty}_{\ell=2} \frac{G}{c^{2\ell+3}} \biggl\{ 2(\ell+1)\mu_{\ell+1} \left( \dot{\bU}_{\!iL} \dot{\bU}_{\!L} + \frac{b_{\ell}b_{\ell+1}}{c^2}\dot{\bV}_{\!iL} \!\dot{\bV}_{\!L}  \right)
\!+ \!\sigma_\ell \,\varepsilon_{ijk}\dot{\bU}_{\!jL-1}
\!\dot{\bV}_{\!kL-1} \biggr\} \label{FluxP}
\end{align}\end{subequations}
where we recall that $b_\ell \equiv \frac{2 \ell}{\ell+1}$ and we defined
\begin{subequations}\begin{align}
\mu_{\ell} & \equiv  \frac{2}{\ell ! \ell!}  \left( m_{\ell-2} - 2m_{\ell-1}+\frac{1}{2}m_{\ell} \right) = \frac{(\ell+1)(\ell+2)}{(\ell-1)\ell \ell!(2\ell+1)!!} , \\
\sigma_\ell & \equiv \frac{4b_\ell}{\ell! \ell!} \left(\frac{m_{\ell -1}}{\ell - 1}- \frac{m_{\ell}}{\ell}\right)=  \frac{8(\ell+2)}{(\ell-1)(\ell+1)!(2\ell+1)!!},
\end{align}\end{subequations}
with $m_\ell$ defined in \eqref{quadint}.
Explicitly, 
\begin{subequations}\label{FluxesEP2}\begin{align}
\dot {\mathcal E} &= - \sum^{+\infty}_{\ell=2} \frac{G}{c^{2\ell+1}}
\biggl\{ \frac{(\ell+1)(\ell+2)}{(\ell-1)\ell \ell!(2\ell+1)!!}
\dot{\bU}_{\!L} \dot{\bU}_{\!L}
+ \frac{4\ell
  (\ell+2)}{c^2(\ell-1)(\ell+1)!(2\ell+1)!!}
\dot{\bV}_{\! L}
\dot{\bV}_{\! L}\biggr\} \,,\label{FluxEe}\\ 
\dot {\mathcal P_i} &=
- \sum^{+\infty}_{\ell=2} \frac{G}{c^{2\ell+3}} \biggl\{
\frac{2(\ell+2)(\ell+3)}{\ell(\ell+1)!(2\ell+3)!!}
\dot{\bU}_{\!iL} \dot{\bU}_{\!L} + \frac{8(\ell+3)}{c^2(\ell+1)!(2\ell+3)!!}
\dot{\bV}_{\!iL}
\!\dot{\bV}_{\!L} \nonumber\\
&\quad \qquad\qquad\qquad+ \frac{8(\ell+2)}{(\ell-1)(\ell+1)!(2\ell+1)!!}  \,\varepsilon_{ijk}
\dot{\bU}_{\!jL-1}
\!\dot{\bV}_{\!kL-1} \biggr\} \,.\label{FluxPe}
\end{align}\end{subequations}
These final expressions agree with Eq.~(4.17) of \cite{Blanchet:2018yqa}, after using Eq.~\eqref{MS}, and with Eq.~(4.14) and Eq.~(4.20) of \cite{Thorne:1980ru}, after using Eq.~\eqref{MS} and averaging over wavelengths. Here, we provided a \emph{covariant} derivation of these flux-balance laws that only relies on the leading radiative behavior of the gravitational field. The result is \emph{exact} at all orders in $c$ and $G$, while the derivations of \cite{Blanchet:2018yqa} used canonical multipole moments in intermediate steps which hinders to prove that the result is in fact exact in terms of radiative multipole moments. While the result of \cite{Thorne:1980ru} required an average over wavelengths, our result shows that this average is not necessary: the right-hand side can be evaluated locally. 

At lowest post-Newtonian order, we can use Eq.~\eqref{MS} to recover the standard formulae \cite{1975ApJ...197..717E,PhysRev.128.2471,Bekenstein:1973zz}
\begin{subequations}
\begin{align}
\dot {\mathcal E} &= - \frac{G}{c^5}\left(\frac{1}{5} I_{ij}^{(3)}I_{ij}^{(3)} \right)- \frac{G}{c^7}\left(\frac{1}{189}I_{ijk}^{(4)}I_{ijk}^{(4)} + \frac{16}{45}J_{ij}^{(3)}J_{ij}^{(3)} \right)+ O\left(c^{-9}\right), \\
\dot {\mathcal P_i} &= - \frac{G}{c^7} \left(\frac{2}{63} I_{ijk}^{(4)}I_{jk}^{(3)} + \frac{16}{45}\eps_{ijk} I_{jl}^{(3)}J_{kl}^{(3)}  \right) + O\left(c^{-9}\right). \label{Pexp}
\end{align}
\end{subequations}
The term of order $G/c^5$ in $\dot{\mathcal E}$ is the Einstein's quadrupole formula.

\paragraph{Angular momentum and center-of-mass balance laws.}
We now turn our attention to the angular momentum and center-of-mass flux-balance laws \eqref{JKflux}. 
The computation consists of substituting the quadratic expressions \eqref{T1dec} and \eqref{T2dec} in the quadratic bilinear operators \eqref{T1T2}, and using the integrals provided in Appendix~\ref{app:int}. We obtain
\begin{subequations}
\begin{align}
 \dot{\mathcal J_i} &= - \varepsilon_{ijk} \sum_{\ell=2}^{+\infty}
\frac{G}{c^{2\ell+1}}\ell\mu_\ell \biggl\{
 {\bU}_{\!jL-1}\dot{\bU}_{\!kL-1} + \frac{b_\ell b_\ell}{c^2} 
{\bV}_{\!jL-1} \dot{\bV}_{\!kL-1}\biggr\} \,,\label{FluxJ0} \\
\dot{ \mathcal K_i} \!+\! u\, \dot{\mathcal P_i}  &= -  \sum^{+\infty}_{\ell=2} \frac{G}{c^{2\ell+3}} (\ell+1)^2\mu_{\ell+1} \! \biggl\{\!
 \dot{\bU}_{\!L}{\bU}_{\!i L} - {\bU}_{\!L}\dot{\bU}_{\!i L} + \frac{b_\ell b_\ell}{c^2} 
\left(\dot{\bV}_{\!L} {\bV}_{\!i L}-{\bV}_{\!L} \dot{\bV}_{\!i L}\right)\!\biggr\}.\label{fluxK200}
\end{align}
\end{subequations}
After the manipulation of the integration coefficients, the Lorentz flux-balance laws read explicitly as
\begin{subequations}
\begin{align}
\dot{\mathcal J_i} &= - \varepsilon_{ijk} \sum^{+\infty}_{\ell=2}
\frac{G}{c^{2\ell+1}} \biggl\{
\frac{(\ell+1)(\ell+2)}{(\ell-1)\ell!(2\ell+1)!!}
{\bU}_{\!jL-1}\dot{\bU}_{\!kL-1}
 \nonumber \\
 & \qquad  \qquad \qquad \qquad+ \frac{4\ell^2
  (\ell+2)}{c^2(\ell-1)(\ell+1)!(2\ell+1)!!}
{\bV}_{\!jL-1}
\dot{\bV}_{\!kL-1}\biggr\} \,,\label{FluxJe}\\
\dot{ \mathcal K_i} + u \, \dot{\mathcal P_i}  & = - \sum^{+\infty}_{\ell=2} \frac{G}{c^{2\ell+3}}
\biggl\{\frac{(\ell+2)(\ell+3)}{\ell\,\ell!(2\ell+3)!!}
( \dot{\bU}_{\!L}{\bU}_{\!i L} - {\bU}_{\!L}\dot{\bU}_{\!i L} )
+\frac{4(\ell+3)}{c^2\ell!(2\ell+3)!!}(
 \dot{\bV}_{\!L}{\bV}_{\!i L} - {\bV}_{\!L}\dot{\bV}_{\!i L} ) \biggr\} \,.\label{FluxG}
\end{align}
\end{subequations}
The result is exact to all orders in $G$ and $c$ in terms of the radiative multipoles. 

As it turns out from the computation above, there are neither parity-odd combinations $\bU_L\bV_{L'}$ nor total $u$-derivatives in the right-hand side of the flux-balance laws for angular momentum and center-of-mass.  It is sometimes stated in the literature that this follows from parity arguments. However, such terms \emph{do} appear in the super-angular momentum flux-balance law; see \eqref{JdotUV} below. Instead, this non-trivial property is rooted in the definition of the Bondi angular momentum aspect \eqref{NA}. Furthermore, if one instead defines the angular momentum using a different prescription such as \eqref{genJ} with $\alpha \neq 1$, the angular momentum flux will be complemented with a parity-odd term, 
\bea\label{fluxJalpha}
\dot{\mathcal J}^{(\alpha)}_i = \dot{\mathcal  J}^{(\alpha = 1)}_i +(\alpha-1) \sum_{\ell = 2}^{+\infty} \frac{G}{c^{2\ell+3}}(\ell+1)^2 \mu_{\ell+1}\frac{d}{du} \left(b_\ell \bU_{iL}\bV_L-b_{\ell+1} \bU_{L}\bV_{iL}\right).
\eea

While the angular momentum flux-balance law \eqref{FluxJe} matches with the final result of \cite{Thorne:1980ru} in its range of validity, our derivation is more general than mentioned in  \cite{Thorne:1980ru}. There, the fluxes are expressed in terms of source multipole moments with the restriction that the source lies in the rest frame and the computation is based upon pseudo-tensors at leading order in the $G$ expansion. Here, the result holds for arbitrary configurations without any restriction (except that Einstein's equations are obeyed!) and the result is exact at all orders in $G$ and $c$ in terms of the radiative multipoles. Such multipoles can be expressed in terms of the source multipoles using Eq.~\eqref{MS}.

Our derivation contrasts with the recent derivation of the angular momentum flux-balance law in \cite{Blanchet:2018yqa}. There, formally divergent (when $r\to +\infty$) terms combines into a total time derivative and vanish after angular integration when the source is at rest, but persist when the source is moving with respect to the asymptotic rest-frame. These persisting total time derivatives amount to a redefinition of the angular momentum. In our derivation, the angular momentum is finite and uniquely defined as \eqref{defJ}-\eqref{NA} and Einstein's equations \eqref{constr2} uniquely determine the finite angular momentum fluxes in terms of the radiative multipole moments as \eqref{FluxJe}.

Our result for the center-of-mass flux-balance law differs from the one proposed by Blanchet and Faye \cite{Blanchet:2018yqa}. After a closer examination, one can match our respective expressions after relating our center-of-mass charges as $\mathcal K_i = \mathcal K^{(BF)}_i + \delta\mathcal K_i$. However, $ \delta\mathcal K_i$ cannot be written as a covariant expression in terms of the metric on the sphere and the shear. Indeed, the only covariant term that one could add to the definition of $N_A$ \eqref{NA}, that does not affect the angular momentum, is a term proportional to $\frac{c^3}{G}\p_A (C_{BC}C^{BC})$. Such a term however does not have the required structure to match $\delta {\mathcal K}_i$. More precisely, we find 
\begin{align}
\dot{\mathcal K}_i^{(\alpha=1,\beta)}&= \dot{\mathcal K}_i^{(\alpha=1,\beta=1)}-(\beta-1)  \sum_{\ell = 2}^\infty \frac{G}{c^{2\ell + 2}}\dfrac{d}{du}\Big[ (\ell+1)\mu_{\ell+1}\big(\bU_{iL} \bU_L + \frac{b_\ell b_{\ell + 1}}{c^2}\bV_{iL}  \bV_L\big)  \nonumber\\
&\hspace{6cm}+\frac{1}{2}\sigma_\ell \eps_{ijk}\bU_{jL-1}\bV_{k L-1}\Big]. 
\end{align}

We conclude that the definition of $\mathcal K^{(BF)}_i $ violates covariance with respect to the metric structure of the celestial sphere, while our definition of $\mathcal K_i$, given by Eq.~\eqref{defK}, is covariant. Instead, the center-of-mass flux \eqref{FluxG} identically agrees at leading and subleading order in the multipolar expansion with Eq.~(31) of \cite{Kozameh:2017qiw} after using their dictionary Eqs.~(41)-(44). 

 At lowest post-Newtonian order, we can use Eq.~\eqref{MS} to obtain the fluxes in terms of the source moments, 
\begin{subequations}
\begin{align}
\dot{ \mathcal J_i} &= - \frac{G}{c^5} \left(\frac{2}{5}\eps_{ijk} I_{jl}^{(2)}I_{kl}^{(3)}\right)- \frac{G}{c^7} \left(\frac{1}{63}\eps_{ijk} I_{jlm}^{(3)}I_{klm}^{(4)} + \frac{32}{45}\eps_{ijk}J_{jl}^{(2)}J_{kl}^{(3)}\right)+ O\left(c^{-9}\right), \\
\dot{ \mathcal G_i}  &= \mathcal  P_i - \frac{G}{c^7}\left[\frac{1}{21}\Big( I_{jk}^{(3)}I_{ijk}^{(3)} - I^{(2)}_{jk} I^{(4)}_{ijk}\Big)\right]+ O\left(c^{-9}\right), 
\end{align}
\end{subequations}
which match with those found by other methods in \cite{1975ApJ...197..717E,Kozameh:2017qiw}.

We can also compare our expression \eqref{FluxG} with the center-of-mass flux obtained in \cite{Nichols:2018qac}. Using the conversion between STF tensors and spherical harmonics as detailed in Appendix \ref{app:sph} we obtain 
\begin{subequations}
\begin{align}
\dot{ \mathcal K_x} \!+ \!u \, \dot{\mathcal P_x}  & \!=\! \frac{1}{64 \pi}  \sum^{+\infty}_{\ell=2}\sum_{m=-\ell}^\ell \frac{G}{c^{2\ell+3}} \text{a}_\ell  \left[ \text{b}^{(+)}_{\ell m} \left(F^{(+)}_{\ell m}-G^{(+)}_{\ell m}\right) - \text{b}^{(-)}_{\ell m} \left(F^{(-)}_{\ell m} - G^{(-)}_{\ell m}\right) \right] ,\label{Kx}\\
\dot{ \mathcal K_y} \!+ \!u \, \dot{\mathcal P_y}  & \!=\!- \frac{i}{64 \pi}  \sum^{+\infty}_{\ell=2}\sum_{m=-\ell}^\ell \frac{G}{c^{2\ell+3}} \text{a}_\ell \left[\text{b}^{(+)}_{\ell m} \left(F^{(+)}_{\ell m}+G^{(+)}_{\ell m}\right)+\text{b}^{(-)}_{\ell m} \left(F^{(-)}_{\ell m}+G^{(-)}_{\ell m}\right)\right] ,\label{Ky}\\
\dot{ \mathcal K_z} \!+ \!u \, \dot{\mathcal P_z}  & \!=\! \frac{1}{32\pi} \sum^{+\infty}_{\ell=2}\sum_{m=-\ell}^\ell \frac{G}{c^{2\ell+3}}\text{a}_\ell \text{c}_{\ell m} \left[ \bar{\text{U}}^{\ell \, m}\dot{\text{U}}^{\ell+1 \, m} \!-\! \left(\bar{\text{U}}^{\ell+1 \, m}\dot{{\text{U}}}^{\ell \, m}\right)\! +\! \frac{1}{c^2}\left(\bar{\text{V}}^{\ell \, m}\dot{ \text{V}}^{\ell+1 \, m}\! -\! \bar{\text{V}}^{\ell+1 \, m}\dot{\text{V}}^{\ell \, m}\right) \right]\label{Kz}
\end{align}
\end{subequations}
where 
\begin{subequations}
\begin{align}
\text{a}_\ell & \equiv \sqrt{\frac{(\ell-1)(\ell+3)}{(2 \ell+1)(2\ell + 3)}}, \\
\text{b}_{\ell m}^{(\pm)} &\equiv \sqrt{(\ell \pm m +1)(\ell \pm m +2)}, \\
\text{c}_{\ell m} &\equiv \sqrt{(\ell+m+1)(\ell-m+1)},\\
F^{(\pm)}_{\ell m}& \equiv  \dot{\text{U}}^{\ell \, m}\bar{\text{U}}^{\ell+1 \, m \pm 1} +\! \frac{1}{c^2}\dot{\text{V}}^{\ell \, m}\bar{ \text{V}}^{\ell+1 \, m \pm 1}\! ,\\
G^{(\pm)}_{\ell m} & \equiv   \bar{\text{U}}^{\ell \, m}\dot{{\text{U}}}^{\ell+1 \, m \pm 1}\! +\! \frac{1}{c^2} \bar{\text{V}}^{\ell \, m}\dot{{\text{V}}}^{\ell+1 \, m \pm 1}.
\end{align}
\end{subequations}
The expressions  \eqref{Kx}-\eqref{Ky}-\eqref{Kz} exactly reproduce Eq.~(2.42) of \cite{Nichols:2018qac} after converting to the convention of fluxes of opposite signs $\dot{ \mathcal K}^{(GW)}_i = -\dot{ \mathcal K_i}$, $\dot{ \mathcal P}^{(GW)}_i = -\dot{ \mathcal P_i}$ (see footnote 7 of \cite{Nichols:2018qac} for the motivation of this sign flip convention).\footnote{For this match, two sign errors were corrected, one in \cite{Nichols:2018qac}, see the upcoming Erratum, and one in the first arXiv version of this paper. We thank D. Nichols for helping obtaining this match.}

\subsection{Supermomentum flux-balance law}
\label{sub:supermomenta}

The general BMS flux-balance laws are obtained by taking the symmetry parameter to be an arbitrary combination of spherical harmonics, \emph{e.g.} $T=T_{L''} N_{L''}$. Note that we use $\ell''$ to label the symmetry parameter as $\ell,\ell'$ are reserved to label the radiative multipoles. So far, we limited ourselves to the lowest $\ell''=0,1$ harmonics for the function $T$ and the vector fields $Y^A$ that generate the Poincar\'e subgroup of supertranslation and super-Lorentz charges \eqref{BMScharge}. In what follows, we will derive the remaining flux-balance laws, starting with the supermomenta.  
We shall use the convention that all supertranslations have the same dimensions as the spatial translations. Indeed, it was shown in \cite{Compere:2016hzt} that, with the exception of the time translation generated by the constant harmonic $\ell'' =0$, all other supertranslations can be understood as spatial transformations in the bulk of spacetime. As a result, all supermomenta will have the same dimensions as the linear momentum.

The flux-balance law of Bondi supermomentum \eqref{fluxPT2} can be expanded in STF harmonics using Eq.~\eqref{BMSsymmSTF}. This gives, schematically,
\bea \label{SmomentaSTFbl}
\dot {\mathcal P}_{T_{L''}} - \left[\dot {\cal P}_{T_{L''}}\right]_{\text{soft}}  = \left[\dot {\cal P}_{T_{L''}}\right]_{\text{hard}}.
\eea

The soft contribution is easily computed. It is non-vanishing only for $\ell'' \geq 2$ and gives
\begin{empheq}[box=\fbox]{align}
\left[\dot {\cal P}_{T_{L''}}\right]_{\text{soft}} &\equiv  - \frac{c^3}{4 G} \oint_S  N_{L''}\boldsymbol{T}_{L''} (\Delta + 2) \Delta \dot C^+ = \frac{\Theta_{\ell'' - 2}}{c^{\ell''-1}} \frac{(\ell''+2)(\ell''+1)}{2 (2\ell'' +1)!! } \boldsymbol{T}_{L''}\dot \bU_{L''},
\end{empheq}
after using Eq.~\eqref{cp}, the property that $\Delta=-\ell(\ell+1)$ when acting on a harmonic function of order $\ell$ and upon integration using \eqref{quadint}. There is no parity-odd contribution (\emph{i.e.}, proportional to $\dot \bV_{L''}$). The soft supertranslation term has a well-understood interpretation. When considering a process which is non-radiative in the far future $u \rightarrow +\infty$ and far past $u \rightarrow -\infty$, the difference of the radiative multipoles $\bU_{L''}|_{u \rightarrow \infty} - \bU_{L''}|_{u \rightarrow -\infty}$ is the displacement memory field \cite{Strominger:2014pwa}, up to an overall normalization. We will further comment about this around Eq.~\eqref{MP}.

In order to compute the hard contribution, we first expand it into radiative multipoles, which is performed in Eq.~\eqref{CC}. We can then use the integration formulae \eqref{supertranslation STF integrals}, detailed in Appendix \ref{app:int}, to obtain the result. 
It is useful to separate the hard contribution in two terms: one contribution consisting of parity-even terms of the form $\bU\bU$ and $\bV\bV$ and denoted with the superscript $+$, and a second contribution consisting of parity-odd terms of the form $\bU\bV$ and denoted with the superscript $-$. In formulae,
\begin{equation}\label{FluxesMl}
\left[\dot {\cal P}_{T_{L''}}\right]_{\text{hard}}\equiv - \frac{c^2}{8G} \oint_S N_{L''}\boldsymbol{T}_{L''} ~  \dot C_{AB} \dot C^{AB}  =\left[\dot {\cal P}_{T_{L''}}\right]_{\text{hard}}^+ + \left[\dot {\cal P}_{T_{L''}}\right]_{\text{hard}}^-.  
\end{equation}
The parity-even contribution is given by
\begin{empheq}[box=\fbox]{align} \label{SupermUU0}
\left[\dot {\cal P}_{T_{L''}}\right]_{\text{hard}}^+ &= -\sum_{\ell,\ell'=2}^\infty \frac{G}{c^{\ell + \ell'+2}} \mu^{P,+}_{\ell,\ell',\ell''} \boldsymbol{T}_{L_1 L_2} \Big( \dot{\bU}_{\! L_1 L_3}  \dot{\bU}_{\! L_2 L_3}+\frac{b_\ell b_{\ell'}}{c^2} \dot{\bV}_{\! L_1 L_3}  \dot{\bV}_{\! L_2 L_3}  \Big)\delta_{\ell,\ell',\ell''}. 
\end{empheq}
The delta symbol $\delta_{\ell,\ell',\ell''}$ is defined in Eq.~\eqref{def delta}, which constrains $\ell_1,\ell_2,\ell_3$ defined in Eq.~\eqref{defell123} to integer values. We can explicitly solve these constraints by taking $|\ell'-\ell|=\ell''-2k$ for $k$ ranging from $0$ to either $\lfloor \frac{\ell''}{2} \rfloor$ or $\lfloor \frac{\ell''-1}{2} \rfloor$; see Eq.~\eqref{eqC21}. The constrained sum over $\ell,\ell'$ can then be replaced by a sum over $\ell,k$ as follows
\begin{align}
\left[\dot {\cal P}_{T_L^{''}}\right]_{\text{hard}}^+ &= -\left(\sum_{k=0}^{\lfloor \frac{\ell''}{2} \rfloor} +\sum_{k=0}^{\lfloor \frac{\ell''-1}{2} \rfloor} \right)  \sum_{\ell=\text{max}(2,k)}^\infty  \frac{G}{c^{2(\ell - k) +2+ \ell''}} \mu^{P,+}_{\ell,\ell+\ell''-2k,\ell''} \nonumber\\
& \qquad\qquad\qquad \quad \times \boldsymbol{T}_{L_1 L_2} \Big( \dot{\bU}_{\! L_1 L_3}  \dot{\bU}_{\! L_2 L_3}+\frac{b_\ell b_{\ell+\ell''-2k}}{c^2} \dot{\bV}_{\! L_1 L_3}  \dot{\bV}_{\! L_2 L_3}  \Big).\label{SupermUU}
\end{align}
Here, $|L_1|=\ell''-k$, $|L_2|=k$ and $|L_3|=\ell - k$. 
The coefficients are given by
\begin{equation}
\mu^{P,+}_{\ell,\ell',\ell''} = \frac{2}{\ell! \ell'!} \left( m_{\ell-2,\ell'-2,\ell''} -2 m_{\ell-1,\ell'-1,\ell''}+\frac{1}{2}m_{\ell,\ell',\ell''}\right).
\end{equation}
For $\ell''=0$, only the first term with $k=0$ is non-vanishing. The coefficient then reads as
\bea         
\mu^{P,+}_{\ell,\ell,0} = \frac{(\ell+1)(\ell+2)}{(\ell-1) \ell \ell! (2\ell+1)!!},
\eea
and it correctly reproduces the coefficient of the energy flux-balance law \eqref{FluxE} with the identification $\mathcal{P}_\emptyset  = E/c$. For $\ell'' = 1$, both terms with $k=0$ add up. The coefficient 
\bea
2\mu^{P,+}_{\ell,\ell+1,1} = \frac{2(\ell+2)(\ell+3)}{\ell(\ell+1)!(2\ell+3)!!}
\eea
reproduces the correct coefficient of the linear-momentum flux-balance law \eqref{FluxP}. 

The parity-odd contribution reads as\footnote{Note that we can freely exchange the upper limit $\lfloor \frac{\ell''-1}{2} \rfloor$ of the first sum with the upper limit $\lfloor \frac{\ell''-2}{2} \rfloor$ of the second sum. They differ only when $\ell''=1+2q$, $q \in \mathbb N$ and in that case $|L_1|=|L_2|$ and the coefficients of each terms are both  $\mu^{P,-}_{\ell,\ell,\ell''}$ and therefore agree. See Appendix \ref{app:int} for a derivation.}
\begin{empheq}[box=\fbox]{align} \label{dPVV}
\left[\dot {\cal P}_{T_L^{''}}\right]_{\text{hard}}^- &=
-  \sum_{k=0}^{\lfloor \frac{\ell''-1}{2} \rfloor} \sum_{\ell=\text{max}(2,k)}^\infty\frac{G }{c^{2(\ell-k) +2+ \ell''}}  \,  \mu^{P,-}_{\ell+\ell''-2k-1,\ell,\ell''}\eps_{imn} \boldsymbol{T}_{iL_1 L_2}
 \dot{\bU}_{\! m L_1 L_3}  \dot{\bV}_{\! n L_2 L_3} \nonumber \\
& \quad - \sum_{k=0}^{\lfloor \frac{\ell''-2}{2} \rfloor}\sum_{\ell=\text{max}(2,k)}^\infty \frac{G }{c^{2(\ell-k)+2+ \ell''}} \,  \mu^{P,-}_{\ell,\ell+\ell''-2k-1,\ell''} \eps_{imn} \boldsymbol{T}_{iL_1 L_2}\dot{\bU}_{\! m L_2  L_3}  \dot{\bV}_{\! n L_1 L_3}.
\end{empheq}
Here, $L'' = iL_1 L_2$ with $|L_1|=\ell''-1 -k$, $|L_2|=k$, $|L_3|=\ell-1-k$. These terms only exists for $\ell'' \geq 1$. The coefficients are 
\bea
\mu^{P,-}_{\ell,\ell',\ell''} = \frac{8 \ell' \ell''}{(\ell'+1)\ell! \ell'!} \left( \frac{m_{\ell-2,\ell'-2,\ell''-1}}{\ell+\ell'+\ell''-2} - \frac{m_{\ell-1,\ell'-1,\ell''-1}}{\ell+\ell'+\ell''}  \right).
\eea
For $\ell''=1$, only the first terms of Eq.~\eqref{dPVV} exists for $k=0$ and  
\bea
\mu^{P,-}_{\ell,\ell,1} = \frac{8(\ell+2)}{(\ell-1) (\ell+1)! (2\ell+1)!!}
\eea 
reproduces the correct coefficient of the momentum \eqref{FluxP}. This provides nontrivial cross-check of our formulae. As a final remark, note also that both coefficients $\mu^{P,\pm}_{\ell,\ell,1}$ are symmetric under $\ell\leftrightarrow\ell'$.

\paragraph{Post-Newtonian analysis}
Let us first discuss the PN order of the parity-even part \eqref{SupermUU}. In our convention, the supermomentum with $\ell'' = 1$ is exactly the linear momentum $\mathcal P_i$ which appears at $3.5$PN order. The PN order of each $\bU \bU$ term in the parity-even contribution is $\ell - k+1+\frac{\ell''}{2} = |L_3| +1+\frac{\ell''}{2} \geq 3$PN. The dominant (lowest) PN term is determined by the maximal number $k$ or, equivalently, by the minimal number of internal contractions $|L_3| \geq 0$. Since $\ell \geq 2$, terms without contractions ($|L_3| =\ell - k = 0$) are realized only for $\ell'' \geq 4$, terms with one contraction ($|L_3| = 1$) are realized for $\ell'' \geq 2$, while two contractions ($|L_3| = 2$) are achieved for any $\ell'' \geq 0$. For the parity-odd contribution \eqref{dPVV}, the PN order of each term is $\ell - k+1+\frac{\ell''}{2} = |L_3| +2+\frac{\ell''}{2}$. Terms without contractions ($|L_3| = 0$) are realized only for $\ell'' \geq 3$, terms with at least one contraction ($|L_3| \geq 1$) are realized for $\ell'' \geq 1$. This leads to the following dominant PN orders for the parity-even term

\begin{center}
\begin{tabular}{|c|c|}
 \hline
 $l''$-pole & leading PN order  \\ 
 \hline
 0, 2, 4 & 3  \\  
 1, 3, 5 & 3.5 \\
 $\ell''\geq 6$ & $\ell''/2 + 1 \geq 4$  \\
 \hline
\end{tabular}
\end{center}
In particular the energy balance law $\ell'' =0$ is $3$PN due to our convention $P_\emptyset = E/c$, but it is restored to  2.5PN order after dropping on each side of the flux-balance law an overall $c$ factor which allows to recognize the left-hand side as the energy flux. Also, the momentum flux-balance law $\ell'' =1$ is $3.5$PN both for the $\bU \bU$ and $\bU \bV$ terms as it should, and one recovers Eq.~\eqref{Pexp}. It is remarkable that the supermomentum flux-balance law for both the quadrupole ($\ell''=2$) and hexadecapole ($\ell''=4$) are 3PN order, which is just intermediate between the energy and the momentum flux-balance laws. In addition, the $\ell''=3,5$ flux-balance laws are just 3.5PN which is of the same order as that of linear momentum. 

Explicitly, the first leading flux-balance laws \eqref{SmomentaSTFbl}, ordered by the leading PN order of their quadratic fluxes up to 3.5PN, read as 
\begin{subequations}\label{Pm}
\begin{align}
\dot{ \mathcal P}_{ij}-\frac{2}{5c}\dot \bU_{ij}  &=  +\frac{G}{c^6}\left[\frac{4}{35}\left(  \dot \bU_{ik}\dot \bU_{jk} -\frac{1}{3}\delta_{ij} \dot \bU_{kl}\dot \bU_{kl}\right)\right] +\mathcal{O}(c^{-8}), \label{Pm1}\\
\dot{ \mathcal P}_{ijkl} -\frac{1}{63c^3}\dot \bU_{ijkl} &=  - \frac{G}{c^6}\left(\frac{2}{315}  \dot \bU_{\langle ij}\dot \bU_{kl \rangle }\right) +\mathcal{O}(c^{-8}),\\
\dot{ \mathcal P}_{ijk}-\frac{2}{21c^2}\dot \bU_{ijk}  &=  +\frac{G}{c^7}\left( \frac{2}{63} \dot \bU_{l \langle ij} \dot \bU_{k\rangle  l} +\frac{8}{105}\eps_{mn \langle i} \dot \bU_{j | m | } \dot \bV_{k \rangle  n}  \right) \!+\! \mathcal{O}(c^{-8}),\label{Pm2} \\
\dot{ \mathcal P}_{ijklm}-\frac{1}{495c^4}\dot \bU_{ijklm} &=  - \frac{G}{c^7}\left(\frac{4}{2079} \dot \bU_{\langle ij}\dot \bU_{klm\rangle }\right)  +\mathcal{O}(c^{-8}).
\end{align}
\end{subequations}
By keeping the soft term in the above equations on the left-hand side and the rest on the right-hand side, one obtains non-trivial equations for the radiative multipoles by rewriting the right-hand side in terms of source multipoles using Eq.~\eqref{MS} and using the dictionary between supermomenta and source multipoles (discussed around Eq.~\eqref{PijB}). In the first equation, the quadrupole-quadrupole interaction responsible for the nonlinear gravitational-wave memory effect arises at 2.5PN, as obtained in \cite{PhysRevD.44.R2945,Arun:2004ff,Favata:2008yd}. Similarly, in the third equation, the octupole radiative mass multipole is sourced by the right-hand side of Eq.~\eqref{Pm2} which arises at 2.5PN. Interestingly, for higher multipoles $\ell''\geq 4$, the quadratic terms associated with non-linear memory arise even earlier at 1.5PN with respect to the corresponding radiative multipole moment.

\subsection{Super-angular momentum flux-balance law}
\label{sec:superrot}

We now compute the super-angular momentum flux-balance laws by expanding Eq.~\eqref{fluxJ2} using the relevant expressions in Appendix \ref{App:T1T2} and \ref{app:int}. The total super-angular momentum flux contains a soft (linear) contribution, as well as a hard (non-linear) contribution, which are both parity-odd. The hard contribution can be split in two sectors: a combination (denoted by $+$) of parity-even quantities of the form $\bU\bU$ and $\bV\bV$ contracted with the Levi-Civita symbol, and another combination (denoted by $-$) of parity-odd quantities of the form $\bU\bV$.
\begin{align}
    \dot {\cal J}_{S_{L''}} - \left[\dot {\cal J}_{S_{L''}}\right]_{\text{soft}} =\left[\dot{\cal J}_{S_{L''}}\right]_{\text{hard}} = \left[\dot{\cal J}_{S_{L''}}\right]^+_{\text{hard}} +\left[\dot{\cal J}_{S_{L''}}\right]^-_{\text{hard}}.
\end{align}
The soft term, appearing on the left-hand side of Eq.~\eqref{fluxJ2}, can be easily computed substituting $C^-$ from Eq.~\eqref{cm}, $\Phi = \frac{1}{\ell''} N_{L''}\bS_{L''}$, using the fact that $\Delta (N_L \boldsymbol{A}_L)=-\ell(\ell+1)(N_L \boldsymbol{A}_L)$, and the integration formula \eqref{quadint} to obtain
\begin{empheq}[box=\fbox]{align}
\left[\dot {\cal J}_{S_{L''}}\right]_{\text{soft}}\equiv u \frac{c^4}{8G}  \oint_S \Delta\Phi   (\Delta + 2) \Delta \dot C^-= \frac{\Theta_{\ell''-2}}{c^{\ell''-1}} \frac{\ell'' (\ell'' +1)(\ell''+2)}{2 (2\ell''+1)!!}u\,\bS_{L''}\dot \bV_{L''}\,.
\end{empheq}
Now we turn to the more involved hard contribution
\begin{align}
\left[\dot{\cal J}_{S_{L''}}\right]_{\text{hard}} \equiv \frac{c^3}{8G}\oint_S \eps^{AB}\pd_B\Phi \left( - 3 T^{(1)}_A(\dot C,C)  + 4 T^{(2)}_A(\dot C,C)  \right).
\end{align}
The $+$ sector of the hard contribution reads as
\begin{empheq}[box=\fbox]{align}\label{J dot UU}
\left[\dot{\cal J}_{S_{L''}}\right]^+_{\text{hard}}&\!=\eps_{kpq}\!\sum_{\ell,\ell'=2}^\infty\dfrac{G}{c^{\ell+\ell'+1}}\, \mu^{J,+}_{\ell,\ell',\ell''}\nonumber\\
&\qquad \qquad \times\bS_{qL_1L_2}\,\Big( \dot{\bU}_{pL_2L_3}\bU_{kL_1L_3}\!+\!\frac{b_{\ell}b_{\ell'}}{c^2}\dot{\bV}_{pL_2L_3}\bV_{kL_1L_3}\Big)\de_{\ell-1,\ell'-1,\ell''-1}
\end{empheq}
where $\ell_{1,2,3}$ are constrained by the delta function at the end of the above expression. Explicitly, $\ell_{1,2,3}=\ell_{1,2,3}(\ell-1,\ell'-1,\ell''-1)$, which means that in Eq.~\eqref{defell123}, $\ell$, $\ell'$ and $\ell''$ are decreased by 1. We don't further expand the sum above as we did in Eq.~\eqref{SupermUU}.
The integration over the 2-sphere leads to the numerical factor
\begin{align}
\mu^{J,+}_{\ell,\ell',\ell''} &=\dfrac{1}{\,\ell!\,\ell'!} \Big\{(\ell'-2)\widehat{m}^+_{\underline{\ell-1},\ell'-1,\ell''-1}-2(\ell'+1)\widehat{m}^-_{\ell-1,\underline{\ell'-1},\ell''-1}\Big\}+ \left(\ell\leftrightarrow \ell'\right)
\end{align}
where the functions $\widehat{m}^+_{\ell,\ell',\ell''}, \widehat{m}^-_{\ell,\ell',\ell''}$ are defined in Eqs.~\eqref{hat m^+}-\eqref{hat m^-} in terms of $\widehat{m}_{\ell,\ell',\ell''}$ in Eq.~\eqref{hat m def}. The $-$ sector of the hard contribution, instead, reads as
\begin{empheq}[box=\fbox]{align}
\left[\dot{\cal J}_{S_{L''}}\right]^-_{\text{hard}}
&=\sum_{\ell,\ell'=2}^\infty\dfrac{G}{c^{\ell+\ell'+2}} b_{\ell'}\,\bS_{ L_1L_2}\Big[\mu^{J,-}_{\ell,\ell',\ell''}\,\big(\bU_{L_2L_3}\dot\bV_{L_1L_3} -  \dot\bU_{L_2L_3}\bV_{L_1L_3}\big) \nonumber\\
&\hspace{4.3cm}+\sigma_{\ell,\ell',\ell''}\dfrac{d}{du}\big(\bU_{L_2L_3}\bV_{L_1L_3}\big)\Big] \de_{\ell,\ell',\ell''}\label{JdotUV}
\end{empheq}
where 
\begin{align}
\mu^{J,-}_{\ell,\ell',\ell''} &=\dfrac{1}{\,\ell!\,\ell'!} \Big\{ (\ell'-2) m_{\ell-2,\ell'-3,\ell''-1} -(\ell+2\ell')m_{\ell-1,\ell'-2,\ell''-1}+(\ell+1)m_{\ell,\ell'-1,\ell''-1}\nonumber\\
&\quad\qquad\qquad-(\ell-2)m_{\ell-2,\ell'-2,\ell''}+3\ell m_{\ell-1,\ell'-1,\ell''}- (\ell+1)m_{\ell,\ell',\ell''}\Big\}+ \left(\ell\leftrightarrow \ell'\right),
\end{align}
and the coefficient of the total time $u$-derivative is given by 
\begin{align}
    \sigma_{\ell,\ell',\ell''}=\dfrac{1}{\,\ell!\,\ell'!}\left(3\alpha_{\ell,\ell',\ell''}-(\ell''-1)\beta_{\ell,\ell',\ell''}\right)
\end{align}
where 
\begin{subequations}
\begin{align}
    \alpha_{\ell,\ell',\ell''} &= \Big\{ (\ell'-2) m_{\ell-2,\ell'-3,\ell''-1} -(\ell+2\ell'-4)m_{\ell-1,\ell'-2,\ell''-1}+(\ell+\ell'-1) m_{\ell,\ell'-1,\ell''-1}\nonumber\\
&\hspace{-1cm}\quad+\ell \left(m_{\ell-2,\ell'-2,\ell''}-m_{\ell-1,\ell'-1,\ell''} \right)\Big\} - \left(\ell\leftrightarrow \ell'\right)\,;\\
	\beta_{\ell,\ell',\ell''}&=m_{\ell-2,\ell'-2,\ell''-2}+2m_{\ell-1,\ell'-1,\ell''-2}+m_{\ell-2,\ell'-2,\ell''}+2m_{\ell-1,\ell'-1,\ell''}+m_{\ell-2,\ell',\ell''-2}\nonumber\\
	&\hspace{-1cm} +m_{\ell,\ell'-2,\ell''-2}-2m_{\ell-1,\ell'-2,\ell''-1}-m_{\ell-2,\ell'-1,\ell''-1}-4m_{\ell-1,\ell',\ell''-1}-2m_{\ell,\ell'-1,\ell''-1}.
\end{align}
\end{subequations}

For the case of rotations, \emph{i.e.}, $\ell''=1$, the flux-balance law for angular momentum $J_i$ (see Eq.~\eqref{FluxJe}, which agrees with \cite{Thorne:1980ru}) is recovered upon taking $S_{q}=\delta_{qi}$.
Note that $\mu^{J,-}_{\ell,\ell+1,1} = 0 = \sigma_{\ell,\ell+1,1}$ and hence the angular momentum flux \eqref{FluxJe} has no mixed $\bU \bV$ term unlike the general super-angular momentum flux.

\paragraph{Post-Newtonian analysis}

Let us derive the leading PN order of each super-angular momentum flux-balance law. The PN order of the $\bU\bU$ contribution is given by $\frac{1}{2}\min (\ell+\ell'+1)$, which can be found by taking into account all the constraints, namely that $\ell,\ell'\geq 2$ and those implied by $\de_{\ell-1,\ell'-1,\ell''-1}$. The PN order of the $\bU\bV$ contribution is instead $\frac{1}{2}\min (\ell+\ell'+2)$ with the constraints that $\ell,\ell'\geq 2$ and those implied by $\de_{\ell,\ell',\ell''}$. We find that, for the leading PN order is given by the following table

\begin{center}
\begin{tabular}{|c|c|}
 \hline
 $l''$-pole & leading PN order \\ 
 \hline
 1, 3 & 2.5  \\  
 2, 4 & 3 \\
 5 & 3.5\\
 $\ell''\geq 6 $ & $\ell''/2 + 1 \geq 4$  \\
 \hline
\end{tabular}
\end{center}
It is remarkable that the octupole ($\ell''=3$) super-angular momentum flux-balance law is at the same PN order as the angular momentum flux-balance law ($\ell''=1$). 
The first few balance laws, ordered by the leading PN order of their quadratic fluxes up to 3.5PN, read as 
\begin{subequations}
\begin{align}
 \dot{\cal J}_{ijk} - \dfrac{2}{7c^2}\,u\,\dot{\bV}_{ijk}&=-\dfrac{G}{c^5}\left(\frac{6}{35}\,\eps_{pq \langle i}\,\dot\bU_{j|p|}\bU_{k \rangle q}\right)\nonumber\\
 &\hspace{-1.5cm} +\frac{G}{c^7}\eps_{pq \langle i}\,\left[-\dfrac{32}{105}\,\dot\bV_{j|p|}\bV_{k \rangle q}+ \dfrac{2}{567}\,\dot\bU_{j|pl|}\bU_{k \rangle ql} +\frac{1}{189}\left(\dot\bU_{|pl|} \bU_{jk\rangle ql} + \dot\bU_{|pl| jk\rangle} \bU_{ql} \right)\right]\nonumber\\
 &\hspace{-1.5cm} + \frac{G}{35 c^7}\left[ \frac{43}{12}\dot\bU_{m\langle k}\bV_{ij\rangle m}-\frac{19}{4}\bU_{m\langle k}\dot\bV_{ij\rangle m}- \frac{106}{27}\bU_{m \langle ij} \dot\bV_{k \rangle m}+\frac{94}{27}\dot\bU_{m \langle ij} \bV_{k \rangle m}\right]\nonumber\\
 &\hspace{-1.5cm} + \mathcal{O}(c^{-8}), \label{l3Sam}
\end{align}
\end{subequations}
\begin{subequations}
\begin{align}
\dot{\cal J}_{ij} - \dfrac{4}{5c}\,u\,\dot{\bV}_{ij}&=+\dfrac{G}{c^6}\,\left[\dfrac{29}{630}\eps_{pq \langle i}\,\left(\dot\bU_{j\rangle pm}\bU_{qm}+\bU_{j \rangle qm}\dot\bU_{pm}\right)-\dfrac{46}{63}\bU_{k \langle i}\dot\bV_{j\rangle k}+\dfrac{202}{315}\dot\bU_{k\langle i}\bV_{j\rangle k}\right]\nonumber\\
&\quad+\mathcal{O}(c^{-8}),\\
\dot{\cal J}_{ijkl} - \dfrac{4}{63c^3}\,u\,\dot{\bV}_{ijkl}&= -\dfrac{G}{c^6}\,\left[\dfrac{11}{378}\eps_{pq \langle i}\,\left(\dot\bU_{|p|l}\bU_{jk \rangle q}+\dot\bU_{|p|jl}\bU_{k \rangle q} \right) +\frac{8}{63}\left( \bU_{\langle ij}\dot\bV_{kl\rangle }+2\dot\bU_{\langle ij}\bV_{jl\rangle}\right)\right]\nonumber\\
&\quad +\mathcal{O}(c^{-8}),\\
\dot{\cal J}_{ijkls} - \dfrac{1}{99c^4}\,u\,\dot{\bV}_{ijkls}&=-\dfrac{G}{c^7}\,\Bigg\{\dfrac{2}{2079}\eps_{pq \langle i}\,\left[5\dot\bU_{|p|jk}\bU_{ls \rangle q}+4\left(\dot\bU_{|p|j}\bU_{kls \rangle q}+\dot\bU_{|p|jkl}\bU_{s \rangle q}\right) \right]\nonumber\\
&\hspace{-2.5cm} +\left[\frac{58}{1155} \dot\bU_{\langle ij}\dot\bV_{kls \rangle} +\dfrac{2}{77}\bU_{\langle ij }\dot\bV_{kls \rangle} +\dfrac{16}{693}  \bU_{\langle ijk}\dot\bV_{ls\rangle} +\dfrac{464}{10395}\dot\bU_{\langle ijk}\bV_{ls \rangle} \right]\Bigg\}+\mathcal{O}(c^{-9}).
\end{align}
\end{subequations}

For the quadrupole, the hard (non-linear) contribution is 2.5PN higher than the soft (linear) contribution, while for $\ell''\geq3$ the hard contribution is only 1.5PN with respect to the soft one.
In particular, our analysis confirms that the leading gravitational-wave flux that generates the spin memory is at 2.5PN order with respect to the super-angular momentum charge and arises for the $\ell = 3$ STF harmonics as analyzed in \cite{Nichols:2017rqr}.

These flux-balance laws remain to be compared with the PN/post-Minkowskian formalism.

\subsection{Super-centre-of-mass flux-balance law}
\label{sec:superboost}

The super-center-of-mass flux-balance law \eqref{sboostlaw} can be expanded in radiative multipoles using the relevant expressions in Appendices \ref{App:T1T2} and \ref{app:int}, along the same line of the previous subsections. In this case, using Eq.~\eqref{BMSsymmSTF} we can write schematically both
\begin{align}
    \dot {\cal K}_{K_{L''}} &= \left[\dot {\cal K}_{K_{L''}}\right]_{\text{soft}} + \left[\dot{\cal K}_{K_{L''}}\right]_{\text{hard}} ,\\ 
\dot{\cal K}_{K_{L''}} +\frac{\ell''+1}{2} u \dot {\cal P}_{K_{L''}}  &=\left[\dot{\cal K}_{K_{L''}}+\frac{\ell''+1}{2} u\dot {\cal P}_{K_{L''}}\right]_{\text{hard}}.
\end{align}
The soft contribution reads as 
\begin{empheq}[box=\fbox]{align} \left[\dot {\cal K}_{K_{L''}}\right]_{\text{soft}}\equiv -\frac{\ell''+1}{2} u \left[\dot {\cal P}_{K_{L''}}\right]_{\text{soft}} = -\frac{\Theta_{\ell'' - 2}}{c^{\ell''-1}} \frac{(\ell''+2)(\ell''+1)^2}{4 (2\ell'' +1)!! } u\boldsymbol{K}_{L''}\dot \bU_{L''},   
\end{empheq}
after using Eq.~\eqref{cp}, the STF decomposition of $\Psi$ in Eq.~\eqref{BMSsymmSTF} and the property that $\Delta=-\ell(\ell+1)$ when acting on a harmonic function of order $\ell$ and upon integration using Eq.~\eqref{quadint}.

The hard contribution to $\dot{\mathcal K}_{K_{L''}}$ involves the hard contribution to the supermomentum which we already computed. Instead, we will simply compute 
\begin{align}
\left[\dot{\cal K}_{K_{L''}}+\frac{\ell''+1}{2} u\dot {\cal P}_{K_{L''}}\right]_{\text{hard}} \equiv - \frac{c^2}{4G} \oint_S \gamma^{AB}\pd_B \Psi ~T^{(1)}_A( \dot C, C).
\end{align}
In order to expand it in radiative multipoles, we substitute the STF decomposition of $\Psi$ and we expand the quadratic operator $T^{(1)}_A$ (defined in Eq.~\eqref{T1T2}) by using Eq.~\eqref{T1dec}. Finally, we use the results of the Appendix \ref{app:int} to perform the integration over the 2-sphere. For the $+$ sector of the hard contribution, we arrive at 
\begin{empheq}[box=\fbox]{align}
\left[\dot{\cal K}_{K_{L''}} +\frac{\ell''+1}{2}  u \dot {\cal P}_{K_{L''}}\right]^+_{\text{hard}}&\!=\!\!\sum_{\ell,\ell'=2}^\infty\dfrac{G}{c^{\ell+\ell'+2}}\,\mu_{\ell,\ell',\ell''}^{K,+}\nonumber\\
& \qquad\times\bK_{L_1L_2}\big( \bU_{L_2L_3} \dot{\bU}_{L_1L_3}\!+\!\dfrac{b_\ell b_{\ell'}}{c^2}\bV_{L_2L_3} \dot{\bV}_{L_1L_3}\big)\de_{\ell,\ell',\ell''}
\end{empheq}
where the discrete delta function $\de_{\ell,\ell',\ell''}$ constraints $\ell_1,\, \ell_2,\ell_3$ given by Eq.~\eqref{defell123}. The coefficient is given by the following expression
\begin{align}
\mu_{\ell,\ell',\ell''}^{K,+}&=\dfrac{1}{\ell!\,\ell'!}\Big\{(\ell'-2)\big({m}^+_{\ell,\ell',\ell''}-{m}^+_{\ell,\ell'-1,\ell''-1}\big)
-2{m}^-_{\ell,\ell'-1,\ell''-1}\Big\}-\left(\ell\leftrightarrow \ell'\right)
\end{align}
where $m^+_{\ell,\ell',\ell''}$ and $m^-_{\ell,\ell',\ell''}$ are defined in Eq.~\eqref{def m^pm} in terms of $m_{\ell,\ell'\,\ell''}$ in Eq.~\eqref{def:m}. 
It is easy to check that the coefficient reduces to the center-of-mass balance law for the special case $\ell''=1$ by noting \eqref{special case m}. 
The $-$ sector of the hard contribution turns out to be
\begin{empheq}[box=\fbox]{align}
\left[\dot{\cal K}_{K_{L''}} + \frac{\ell''+1}{2} u \dot {\cal P}_{K_{L''}}\right]^-_{\text{hard}}\!&=\!\sum_{\ell,\ell'=2}^\infty\dfrac{G}{c^{\ell+\ell'+3}}\,b_\ell\,\mu_{\ell,\ell',\ell''}^{K,-}\nonumber\\
&\hspace{-1cm}\times\eps_{kpq}\bK_{qL_1L_2}\left( \dot{\bU}_{pL_1L_3}\bV_{kL_2L_3}-{\bU}_{pL_1L_3}\dot\bV_{kL_2L_3}\right)\de_{\ell-1,\ell'-1,\ell''-1}
\end{empheq}
where $\ell_{1,2,3}=\ell_{1,2,3}(\ell-1,\ell'-1,\ell''-1)$ and 
\begin{align}
\mu_{\ell,\ell',\ell''}^{K,-}&=\,\dfrac{1}{\ell!\ell'!}\Big\{\widehat{m}_{\underline{\ell-2},\ell'-2,\ell''-1}-\widehat{m}_{\underline{\ell-1},\ell'-1,\ell''-1}-\ell\big(\widehat{m}_{\ell-2,\ell'-2,\underline{\ell''-1}}-\widehat{m}_{\ell-1,\ell'-1,\underline{\ell''-1}}\big)\nonumber\\
& \qquad\qquad+(\ell-2)\widehat{m}_{\ell-3,\ell'-2,\underline{\ell''-2}}-(\ell-3)\widehat{m}_{\ell-2,\ell'-1,\underline{\ell''-2}}\Big\}- \left(\ell\leftrightarrow\ell'\right).
\end{align}
For $\ell''=1$ the last two terms in the second line vanish by definition of $\widehat{m}_{\ell,\ell',\ell''}$, while it is easy to check using Eq.~\eqref{m hat l''=0} that the remaining terms in the first line add up to zero. Thus, we recover the result that the flux-balance associated with Lorentz boosts does not display any mixed term $\bU \bV$; see Eq.~\eqref{FluxG} above.

\paragraph{Post-Newtonian analysis} 
Now we turn to the PN analysis of most leading super-center of mass flux-balance laws. The analysis is computationally similar to the previous case of super-angular momentum flux-balance law, so we omit the details here.
We find that the PN order of the flux of the superboost charge of rank $\ell''$ reads
\begin{center}
\begin{tabular}{|c|c|}
 \hline
 $l''$-pole & leading PN order \\ 
 \hline
 1, 3, 5 & 3.5  \\  
 2, 4, 6 & 4 \\
 $\ell''\geq 7 $ & $\ell''/2 + 1 \geq 4.5$  \\
 \hline
\end{tabular}
\end{center}
The explicit expressions for the most leading balance laws,  ordered by the PN order of their fluxes up to 4PN order, read as
\begin{subequations}
\begin{align}
\dot{\cal K}_{ijk}+\frac{4}{21c^2}u~\dot{\bU}_{ijk}&=+\dfrac{G}{c^7}\left[\frac{1}{63}\left(\dot\bU_{m \langle ij}\bU_{k \rangle m}-\dot\bU_{m \langle i}\bU_{jk \rangle m}\right)\right]+\mathcal{O}(c^{-9}),\label{KijkF}\\
\dot{\cal K}_{ijkls} + \frac{1}{165c^4}u~\dot{\bU}_{ijkls}&=-\dfrac{G}{c^7}\left[\frac{2}{3465}\,\left(\dot\bU_{\langle ijk}\bU_{ls \rangle}-\bU_{\langle ijk}\dot\bU_{ls \rangle}\right)\right]+\mathcal{O}(c^{-9})\,,\label{KijklsF}\\ 
\dot{\cal K}_{ij} + \frac{3}{5c}u~\dot{\bU}_{ij}&=+\dfrac{G}{c^8}\Bigg\{\left[-\dfrac{1}{216}\dot\bU_{ijmn}\bU_{mn}+\dfrac{1}{42}\,\eps_{pq\langle i}\,\left(\dfrac{4}{3}\dot\bU_{j\rangle pm}\bV_{qm}-\dfrac{3}{2}\bV_{j\rangle qm}\dot\bU_{pm}\right)\right] \nonumber\\
     &\hspace{2.1cm}-\text{dot inverted}\Bigg\} +\mathcal{O}(c^{-9}),\\
\dot{\cal K}_{ijkl}+\frac{5}{126c^3}u~\dot{\bU}_{ijkl}&=+\dfrac{G}{ c^8}\Bigg\{\left[\dfrac{1}{330}\dot\bU_{m\langle ijk}\bU_{l\rangle m}-\dfrac{1}{315}\,\eps_{pq \langle i}\,\left(\dfrac{4}{3}\dot\bU_{|p|jk}\bV_{l \rangle q}-\dfrac{3}{2}\dot\bU_{|p|l}\bV_{jk \rangle q}\right)\right]  \nonumber\\
     &\hspace{2.1cm}-\text{dot inverted}\Bigg\} +\mathcal{O}(c^{-9}).
\end{align}
\end{subequations}
These flux-balance laws remain to be compared with the PN/post-Minkowskian formalism\footnote{Also note that the right-hand side of Eqs.~\eqref{KijkF}-\eqref{KijklsF} can be compared with Eqs.~(4.33a)-(4.33b) of \cite{Nichols:2018qac}. Up to a positive overall factor, our Eq.~\eqref{KijkF} and his Eq.~(4.33a) agree while there is a sign mismatch between our Eq.~\eqref{KijklsF} and Eq.(4.33b). That sign mismatch is resolved by a correction in \cite{Nichols:2018qac} (Nichols, private communication).}

\subsection{BMS flux-balance laws as constraints on source evolution}

Let us consider the gravitational radiation emitted by a compact binary merger. It is well-known, since the seminal papers by Peters and Mathews \cite{PhysRev.131.435,1964PhRv..136.1224P}, that at the lowest PN order, the energy and angular momentum flux-balance laws can be used as evolution equations to compute the secular change of the major axis and the eccentricity of compact binary systems. More fundamentally, the energy and angular momentum flux-balance laws inform upon the radiation-reaction force of the source which starts at 2.5PN order beyond Keplerian motion \cite{1969ApJ...158..997T,1969ApJ...158...45C,1969ApJ...158...55C,1970ApJ...160..153C,1975GReGr...6..197A,1980NYASA.336..279E,1980GReGr..12..467K,1980GReGr..12..521K,1981GReGr..13..335P,1981GReGr..13..777B,1981PhLA...87...81D,1982GReGr..14..181B,damour:1981bb,damour:1982bb,Damour:1983aa,1985AnPhy.161...81S,Iyer:1993xi,Iyer:1995rn}.  

The post-Newtonian/post-Minkowskian formalism (see \emph{e.g.}, \cite{1976ApJ...210..764W,Thorne:1980ru,Blanchet:1985sp,Will:1996zj}) allows to relate the radiative multipoles to the source parameters $\{ p_i \}$, such as the binary masses, relative distance, angular velocity, spins, finite size parameters, \emph{etc}. The number of such parameters can be infinite once one includes all the multipole structure of the sources, e.g. neutron stars with specific internal dynamics. While this is beyond the scope of this paper, one could find in principle the coordinate transformation between Bondi gauge and de Donder gauge perturbatively in $G$, in order to find the map between, on the one hand, the Bondi data $m,\, N_A$ and $C_{AB}$ and, on the other hand, the canonical multipole moments defined in de Donder gauge. As an illustration, by consistency between \eqref{Pm1} and Eq.~(88) of \cite{Blanchet:2013haa}, we can infer the expression of the Bondi quadrupole supermomentum in terms of the canonical multipole moments, which involves retarded integrals,
\begin{align}
\mathcal P_{ij} &= +\frac{2}{5c}\stackrel{(2)}{M}_{ij} + \frac{4G M}{5c^4}\int^\infty_{0} d\tau \left[ \text{ln}\left( \frac{c \tau}{2 r_0} \right) + \frac{11}{12} \right]\stackrel{(4)}{M}_{ij}(u-\tau)\nonumber\\
&\quad+\frac{2G}{5c^6}\left[\frac{1}{7}\stackrel{(5)}{M}_{k \langle i}M_{j \rangle k}-\frac{5}{7}\stackrel{(4)}{M}_{k \langle i}\stackrel{(1)}{M}_{j \rangle k}-\frac{2}{7}\stackrel{(3)}{M}_{k \langle i}\stackrel{(2)}{M}_{j \rangle k}+\frac{1}{3}\eps_{kl \langle i}\stackrel{(4)}{M}_{j \rangle k}S_l \right]\nonumber\\
& \quad+\mathcal{O}\left(\frac{G^2}{c^7}\right) +\mathcal{O}\left(\frac{G}{c^8}\right).\label{PijB} 
\end{align}

In turn, the canonical multipole moments can be expressed in terms of the source multipole moments, which can be themselves expressed in terms of the source parameters $\{ p_i \}$ in a PN expansion. Therefore, such a map defines the functions $m(\{ p_i \})$ and $N_A(\{ p_i \})$, up to residual gauge choices (choice of supertranslation and Lorentz frame, choice of de Donder frame and choice of source coordinates).

The BMS flux-balance laws \eqref{constr1}-\eqref{constr2}-\eqref{NA} or, equivalently, Eqs.~\eqref{BMSfluxes} can then be rewritten as consistency constraints on the  evolution of the source parameters in a fixed residual gauge choice, \vspace{-0.0cm}
\begin{subequations} \label{dotpi}
\begin{align}
\sum_i \frac{\p m}{\p p_i} \dot p_i &= F_u(\{ p_j \}), \\
\sum_i \frac{\p N_A}{\p p_i} \dot p_i &= F_A(\{ p_j \})
\end{align}
\end{subequations}
where the fluxes of $m$ and $N_A$ (including the soft/memory terms) are written as $F_u$ and $F_A$ in terms of the source parameters in post-Newtonian/post-Minkowskian expansions. These constraints are coupled non-linear integro-differential equations, which are equivalent to a subset of Einstein's equations that have already been partially solved. Under the assumption of no incoming radiation from past null infinity, these equations depend upon retarded integrals. By construction, these equations only inform about the radiation-reaction forces. In the small velocity approximation, retarded potentials can be expanded in terms of Newtonian instantaneous expressions and at least at the lowest 2.5PN order, the integro-differential equations can be rewritten as ordinary differential equations; see a related discussion in \cite{Damour:1983aa}. At lowest 2.5PN order, the energy and angular momentum flux balance laws lead to the Peters and Mathews differential equations \cite{PhysRev.131.435,1964PhRv..136.1224P}. We saw earlier that the octupole super-angular momentum flux-balance law \eqref{l3Sam} also arises at 2.5PN order but it has not been used so far for constraining sources. At 4PN order, the tails such as the one appearing in Eq.~\eqref{PijB} introduce a nonlocal-in-time dynamics of the sources \cite{Blanchet:1987wq}.  We leave the derivation of the explicit form of Eqs.~\eqref{dotpi} for generic BMS flux-balance laws for future work.

\section{Global conservation laws for binary mergers}
\label{sec:glob}
The BMS flux-balance laws describe the evolution of the Bondi mass and angular momentum aspects in any spherical direction and at any retarded time. For gravitational systems evolving from a non-radiative state at early retarded time to a non-radiative state at late retarded time, the BMS flux-balance laws can be integrated to relate the difference between initial and final BMS charges to the total gravitational and electromagnetic radiation \cite{Payne:1982aa,Strominger:2014pwa,Flanagan:2015pxa,Compere:2016hzt,Hawking:2016sgy,Bonga:2018gzr,Ashtekar:2019viz}. We will summarize these global constraints for each of the BMS flux-balance laws and derive in particular the initial and final BMS charges for binary black hole mergers. The specification of both the initial and final BMS charges requires a choice of supertranslation and Lorentz frame, \emph{i.e.} a fixation of the BMS asymptotic symmetry group.

\subsection{Conservation of Poincar\'e charges and proper BMS memories}

Qualitatively, we need to distinguish the Poincar\'e conservation laws and the proper BMS conservation laws. The Poincar\'e conservation laws are the global conservation of energy-momentum, angular momentum and center-of-mass. These conservation laws allow, given the data of an initial binary system and given the data of the fluxes of radiation, to deduce the final energy, the final momentum (also called the velocity kick), the final angular momentum and the final center-of-mass (also called the center-of-mass shift). These ten global conservation laws take the form 
\bea
Q|_{u_2} - Q|_{u_1} = \text{Fluxes}\label{Fluxes}
\eea
where $u_1$ and $u_2$ are the initial and final non-radiative final states, the charges $Q$ are the ten Poincar\'e charges \eqref{convPi}-\eqref{defJ}-\eqref{defK} (with $T$, $\Phi$, $\Psi$ a combination of $\ell =0,1$ harmonics) and the fluxes are detailed in Eqs.~\eqref{FluxEe}-\eqref{FluxPe}-\eqref{FluxJe}-\eqref{FluxG}. We will derive explicit expressions for the Poincar\`e charges of an initial infinitely-separated black hole binary and of a final black hole, \emph{i.e.} the left-hand side of \eqref{Fluxes}. This is a nontrivial task since the Poincar\'e charges depend upon the choice of supertranslation and Lorentz frames.

The (infinite set of) proper BMS global conservation laws  have the qualitatively distinct form
\bea
Q|_{u_2} - Q|_{u_1} -\text{Fluxes} =  \text{Memory} \label{flm}
\eea
where $u_1$ and $u_2$ are the initial and final non-radiative final states, the charges $Q$ are the proper BMS charges \eqref{convPi}-\eqref{defJ}-\eqref{defK}, whose fluxes are all terms in  Eqs.~\eqref{fluxPT2}-\eqref{fluxJ2}-\eqref{sboostlaw} either proportional to the matter stress-tensor or quadratic in the news $\dot C_{AB}$ or shear $C_{AB}$ and the memory term are all terms in  Eqs.~\eqref{fluxPT2}-\eqref{fluxJ2}-\eqref{sboostlaw} linear in the news or shear. More precisely, we define the displacement memory $M_{\mathcal P}$, the spin memory $M_{\mathcal J}$ \cite{Pasterski:2015tva,Nichols:2017rqr,Himwich:2019qmj} and the center-of-mass memory $M_{\mathcal K}$ \cite{Nichols:2018qac} as
\begin{subequations}
\begin{align}
M_{\mathcal P} &= -\frac{c^3}{4G}\Delta (\Delta +2) \int_{u_1}^{u_2} du ~\partial_u C^+  = -\frac{c^3}{4G}\Delta (\Delta +2) \left[ C^+ \right]^{u=u_2}_{u=u_1} ,\label{MP} \\
M_{\mathcal J} &= -\frac{c^4}{8G} \Delta^2 (\Delta +2) \int_{u_1}^{u_2} du~ u \p_u C^- ,\label{MJ} \\
M_{\mathcal K} &= +\frac{c^3}{8G}  \Delta^2 (\Delta +2) \int_{u_1}^{u_2} du~ u \p_u C^+ .\label{MK}
\end{align}
\end{subequations}
The transformation law of $C^+$ under supertranslations and Lorentz transformations can be found in Eq. (3.22) of \cite{Compere:2018ylh}, $\delta_{T,Y}C^+ = T + Y^A D_A C^+-\frac{1}{2}C^+ D_A Y^A$, while $C^-$ is invariant under supertranslations. By construction, the memories are supertranslation-invariant observables since such transformations equally shift the initial and final $C^+$. The operators $\Delta (\Delta +2)$ and $\Delta^2(\Delta +2)$ admit as a kernel the lowest 4 spherical harmonics. These operators discard the arbitrary lowest 4 harmonics of $C^\pm$ that do not appear in the metric. For all higher harmonics these operators are invertible. 

The left-hand side of Eq.~\eqref{flm} is therefore uniquely determined by the physical parameters of the initial binary system and the final stationary state. However, the charge difference $Q|_{u_2} - Q|_{u_1}$ and the fluxes also individually dependent upon the choice of supertranslation and Lorentz frames. The transformation laws of the charges and fluxes can be found, \emph{e.g.}, in \cite{Barnich:2011mi,Barnich:2013axa,Barnich:2016lyg,Nichols:2018qac}. The proper BMS global conservation laws can therefore be used to provide the values of the memory fields as a function of the proper BMS charges of the initial and final states in a given supertranslation and Lorentz frame, and as a function of the radiation fluxes in the same frame. We will provide in the following the initial and final BMS charges for binary black hole mergers in an arbitrary supertranslation and Lorentz frame. 

\subsection{The final boosted and supertranslated Kerr metric}

The final state of collapse in General Relativity is described by the Kerr metric, up to a diffeomorphism. In Bondi gauge, the residual diffeomorphisms form the extended BMS group and are associated with nontrivial surface charges, as we reviewed in Section \ref{sec:Bondi}. While all supertranslations preserve asymptotic flatness, only the Lorentz subgroup of the super-Lorentz group preserve asymptotic flatness. In this paper, we will only consider physical processes that preserve asymptotic flatness and we therefore discard ``cosmic'' transitions that induce super-Lorentz transformations \cite{Strominger:2016wns,Compere:2018ylh}. The question that we need to answer is: what is the value of the Bondi mass aspect $m$ and angular momentum aspect $N_A$ for the Kerr metric in Bondi gauge in an arbitrary supertranslation frame and Lorentz frame?

 It is well known that the Bondi mass aspect $m$, whose definition is universally accepted, is invariant under supertranslations and rotations. In any given boosted frame determined by the velocity $\vec{v}$, it is given by \cite{1962RSPSA.269...21B}
\bea
m = \frac{m_{rest}}{\gamma^3 \left(1- \dfrac{\vec{v}}{c} \cdot \vec{n}\right)^3},\qquad \gamma(v) = \frac{1}{\sqrt{1-\dfrac{v^2}{c^2}}},\qquad \vec{n}\cdot \vec{n}=1
\label{mfinal}
\eea
where $\vec{n} = (\sin\theta \cos\phi,\sin\theta\sin\phi,\cos\theta)$ is the unit vector and $m_{rest}$ is the rest mass of the system. We will rederive this expression in Appendix \ref{app:Kerr}. The zeroth moment of the Bondi mass aspect is the energy.\footnote{Remember that the Bondi ``mass'' $m$ has dimension of energy.} The boosted energy $\oint_S m =\gamma m_{rest} $  agrees with the standard special relativistic expression. The dipole moment is the momentum and the expression again agrees with the relativistic expression, $\oint_S  m n_i =\gamma m_{rest} v_i$. The higher moments (\emph{i.e.}, the supermomenta) are specific to General Relativity. For instance, the quadrupole reads as 
\begin{align}
\mathcal P_{ij} = \frac{1}{c} \oint_S  m \, n_{\langle i}n_{j \rangle} &= \frac{3 c^2 m_{rest}}{2 \gamma^3 v^3 }\left[\text{arctanh}\left(\frac{v}{c}\right) - \frac{\gamma^4 v}{c}\left(1-\frac{5}{3}\frac{v^2}{c^2}\right)\right]\left(\frac{v_i v_j}{v^2}-\frac{1}{3}\delta_{ij}\right) \nonumber \\
&=\frac{4 m_{rest}}{5c} \left(\frac{v_i v_j}{c^2}- \frac{v^2}{3c^2}\delta_{ij}\right)+O(c^{-5}). \label{Pijexpl}
\end{align}

The angular momentum aspect $N_A$ in such an arbitrary frame has not yet been derived in closed form, though it is implicit in the literature (see in particular \cite{Ashtekar:1981bq,Barnich:2016lyg}).  The angular momentum aspect $\bar N_A$ can be identified directly from the $1/r$ term of $g_{uA}$ in the Bondi metric expansion \eqref{metricBondi}. This is the quantity that we need to compute. This quantity depends upon the supertranslation frame, the Kerr energy $Mc^2$ and angular momentum $J= M a \, c $ (where $a$ has dimension of length), the angles $\theta,\phi$ and the Lorentz boost parameter $\vec{v}$. 

In an arbitrary supertranslation frame at rest, the Bondi shear is given in terms of the supertranslation field $C(\theta,\phi)$ as \cite{1962RSPSA.269...21B,Strominger:2013jfa}
\bea
C_{AB}=-2 D_A D_B C + \gamma_{AB} \,\Delta C.\label{CABfinal}
\eea
In other words, in the decomposition \eqref{Cpm}, $C^+ = C(\theta,\phi)$ while $C^- = 0$ (it cannot be generated by a diffeomorphism). A related statement is that, out of the two polarizations of the graviton, only one combination of the polarizations exists in the soft limit \cite{Strominger:2014pwa}. For the Schwarzschild black hole equipped with the supertranslation field and at rest, the angular momentum aspect $\bar N_A$ is \cite{Compere:2016jwb,Compere:2016hzt}
\bea
\bar N_A = 3 M c \, \partial_A C. 
\eea
For the Kerr black hole, we have $\bar N_A = - 3 M a\, c \sin^2\theta \p_A \phi$ and the total angular momentum associated with $-\p_\phi$ is indeed $J= M a\, c$. For a Kerr black hole equipped with the supertranslation field and at rest, the angular momentum aspect is \cite{Flanagan:2015pxa} (see their Eq.~(3.17)),
\bea
\bar N_A = 3 M c \, \partial_A C- 3 M a\, c \sin^2\theta \p_A \phi. 
\eea
We now need to consider a finite boost of velocity $v_i$. As shown in \cite{Compere:2018ylh}, the Bondi shear is now given by 
\bea
C_{AB}=\left(u+\frac{C}{c}\right) N_{AB}^{\text{vac}}-2 D_A D_B C + \gamma_{AB} \,\Delta C\label{CABfinalboosted}
\eea
where $N_{AB}^{\text{vac}}=[\frac{1}{2}D_A \Phi D_B \Phi - D_A D_B \Phi]^{TF}$. 
Here TF denotes the trace-free part, $\Phi=0$, and the field $C$ contains a contribution $-n_i v_i$.
The final result for $\bar N_A$, obtained after a quite long computation outlined in Appendix \ref{app:Kerr}, reads as 
\begin{empheq}[box=\widefbox]{align}
\bar N_A =\dfrac{2 m}{c}\p_A C + \p_A \left[m\left(u+\frac{C}{c}\right) \right]  - \frac{3 J }{c^2 \gamma^2 \left(1-\dfrac{\vec{v}}{c} \cdot \vec{n}\right)^2 } \sin^2 \theta' \p_A \phi' .\label{NAfinal}
\end{empheq}
Here, the Bondi mass aspect $m$ is defined in Eq.~\eqref{mfinal}, where $m_{rest}=Mc^2$ is the energy of Kerr at rest and $J$ is the angular momentum at rest or intrinsic spin. The angles in the Lorentz-transformed frame are denoted as $(\theta',\phi')$; see Eq.~\eqref{thetap}. The Kerr metric in an arbitrary Lorentz and supertranslation frame labelled by $\vec{v}$ and $C(\theta,\phi)$, and an arbitrary rotation is finally given by Eq.~\eqref{metricBondi} with $m,\bar N_A$ and $C_{AB}$ defined as in Eqs.~\eqref{mfinal}, \eqref{NAfinal} and \eqref{CABfinalboosted}. Note that the final expression does not contain terms quadratic in the shear. This property is not obeyed by alternative definitions of Bondi angular momentum aspect, including Eq.~\eqref{NA}, which differ by quadratic terms in the shear $C_{AB}$. Explicitly,
\bea
N_A \!=\!- \frac{3 J \sin^2 \theta' \p_A \phi'}{c^2 \gamma^2 \left(\! 1 \! -\! \dfrac{\vec{v}}{c} \cdot \vec{n}\! \right)^{\!2} } \! +\!  \frac{3m \p_A C}{ c}  \! + \!\frac{C \p_A m }{c} \! -\!\frac{c^3}{4G} C_{AB}D_C C^{BC} \!-\! \frac{c^3}{16 G} \p_A (C_{BC} C^{BC}). \label{NAstat}
\eea
In the last stage of this work, we noticed \cite{Ashtekar:2019rpv} where a related expression is derived in another formalism, which remains to be compared with Eq.~\eqref{NAstat}. 

Let us comment on some physics that can be deduced from the expression \eqref{NAfinal}. First remember that under a supertranslation $T(\theta,\phi)$, the supertranslation field changes as $C \mapsto C + T$ while for non-radiative configurations the Bondi angular momentum aspect changes as
\bea
\bar N_A \mapsto \bar N_A +3 \frac{m}{ c}  D_A T + \frac{T}{c} \p_A m ,\label{tran2}
\eea
as can be deduced from Eq.~\eqref{NAfinal} or, \emph{e.g.}, Eq.~(2.24) of \cite{Compere:2018ylh}. The first $\ell=0,1$ harmonics of $C$ do not contribute to the shear $C_{AB}$ and can be interpreted as reference spacetime position $\bar X^\mu$; see Eq.~\eqref{dictCp}. The $\ell \geq 2$ multipoles are specific to General Relativity. The expression \eqref{NAfinal} suggests to define the \emph{intrinsic angular momentum aspect} as 
\bea
N_A^{\text{(intrinsic)}}(\bar X) \equiv \bar N_A - \frac{2 m}{c}\p_A C|_{\ell \geq 2} -\p_A \left[m\left(u+\frac{C|_{\ell \geq 2}}{c}\right) \right]
\eea
and the \emph{intrinsic angular momentum and center-of-mass} as 
\begin{subequations}
\begin{empheq}[box=\widefbox]{align}
\mathcal J_\Phi^{(intrinsic)} & \equiv \oint_S \frac{1}{2} \eps^{AB}\p_B \Phi N_A^{\text{(intrinsic)}}, \\
\mathcal G_\Psi^{(intrinsic)} & \equiv \oint_S \frac{1}{2} \gamma^{AB}\p_B \Psi N_A^{\text{(intrinsic)}}.
\end{empheq}
\end{subequations}
For the Kerr black hole, 
\bea
N_A^{\text{(intrinsic)}} =3 \frac{m}{ c}  D_A (\bar X^i n_i) + \frac{\bar X^0}{c} \p_A m - \frac{3 J }{c^2 \gamma^2 \left(1-\dfrac{\vec{v}}{c} \cdot \vec{n}\right)^2 } \sin^2 \theta' \p_A \phi'.
\eea

The intrinsic angular momentum is free from supertranslation ambiguities for non-radiative configurations and it only transforms under the Poincar\'e group. In particular the global angular momentum $\mathcal J_{n_z}$ is equal to $J$ where $n_z = \p_{\phi'}$. The intrinsic angular momentum aspect is defined as a non-local functional of the metric fields in Bondi gauge, since $C$ is non-local. It provides an explicit expression of the supertranslation-free definition of angular momentum obtained from other methods in \cite{Javadinazhed:2018mle}. It would be interesting to generalize our definition for radiating configurations.

\subsection{The initial binary system of Kerr black holes}
\label{sec:init}

We consider as initial system at $u \rightarrow -\infty$ two Kerr black holes of respective rest mass, rest spin, position and velocity with respect to the frame given by $(m_1,\vec{J_1},\vec{x_1},\vec{v_1})$ and $(m_2,\vec{J_2},\vec{x_2},\vec{v_2})$. At $ u \rightarrow -\infty$, we take a spatial distance $L = |\vec{x_2} - \vec{x_1}| \rightarrow \infty$. Since the binding energy decreases as $O(1/L)$, the total Bondi mass aspect of the system is given by the sum of the two individual Bondi mass aspects. Using Eq.~\eqref{mfinal}, we have 
\begin{empheq}[box=\widefbox]{align}\label{empheq}
m |_\text{init}  \equiv \dfrac{m_1}{\gamma_1^3\left(1-\dfrac{\vec{v}_1 \cdot \vec{n}}{c}\right)^3}+ \dfrac{m_2 }{\gamma_2^3\left(1-\dfrac{\vec{v}_2 \cdot \vec{n}}{c}\right)^3}  
\end{empheq}
where $\gamma_i=\gamma(v_i)$. We will fix the initial supertranslation frame by setting $C = C  |_\text{init}  (\theta,\phi)$ arbitrary and we define $C_{AB}\big|_\text{init} = -2D_A D_B C\big|_\text{init} +\gamma_{AB}\Delta C\big|_\text{init}$. As a consistency check, we can compare the expression for the Bondi supermomentum quadrupole $\mathcal P_{ij}$, as obtained from Eq.~\eqref{empheq}, which is the sum of two terms of the form \eqref{Pijexpl}, and the  expression \eqref{PijB} evaluated at $u \rightarrow -\infty$. After using $\stackrel{(2)}{M}_{ij}=2 \dfrac{m_1}{c^2}v^1_i v^1_j + 2\dfrac{m_2}{c^2}v^2_i v^2_j$, the expressions match at lowest PN order.

We now note that the angular momentum aspect $\bar N_A$ as defined in Eq.~\eqref{metricBondi} leads to an expression for a single black hole \eqref{NAfinal} which contains two parts: a quadratic part of the form $m\, C$ and a part linear in $m$ or $J$. By linearity, the binary system will have the part of the angular momentum aspect linear in $m$ or $J$ given by the linear sum of the two individual bodies up to $O(L^{-1})$ corrections that vanish in the limit $u \rightarrow -\infty$. The quadratic part in $m\, C$ will be given by the total Bondi mass and supertranslation frame, consistently with the transformation law \eqref{tran2}. Therefore, the total angular momentum aspect for the initial binary is
\begin{empheq}[box=\fbox]{align}
\bar N_{A} \big|_\text{init} & \equiv 
\frac{2 m|_\text{init}}{c}\p_A C |_\text{init} + \p_A \left[m |_\text{init}\left(u+\frac{C |_\text{init}}{c}\right) \right]\nonumber \\ 
 &\quad  - \frac{3 J_1  \sin^2 \theta_1 \p_A \phi_1 }{c^2  (\gamma_1)^2 \left(1-\dfrac{\vec{v_1}}{c} \cdot \vec{n}\right)^2 } - \dfrac{3 J_2  \sin^2 \theta_2 \p_A \phi_2 }{c^2  (\gamma_2)^2 \left(1-\dfrac{\vec{v_2}}{c} \cdot \vec{n}\right)^2 }\label{NAex}
\end{empheq}
where the angles $\theta_{1,2},\phi_{1,2}$ are obtained from $(\theta,\phi)$ by a rotation and boost \eqref{thetap} adjusted to each intrinsic spin direction $\vec{J}_{1,2}$ and velocity $\vec{v}_{1,2}$. The spin magnitudes are denoted as $J_1 = \sqrt{\vec{J}_1 \cdot \vec{J}_1}$, $J_2 = \sqrt{\vec{J}_2 \cdot \vec{J}_2}$. The initial Bondi angular momentum aspect \eqref{NA} is finally given by
\bea
N_A|_{init}=\bar N_{A} \big|_\text{init}- u \p_A m\big|_\text{init} -\frac{c^3}{4G} C_{AB} D_C C^{BC}\big|_\text{init} -\frac{c^3}{16G} \partial_A \left(C_{BC} C^{BC}\right) \big|_\text{init}.
\eea

\section{Conclusion}
\label{sec:ccl}

We obtained a simplified form for all BMS (supermomentum, super-angular momentum and super-center-of-mass) flux-balance laws that are obtained from Einstein's constraint equations, which extends earlier work \cite{Payne:1982aa,Barnich:2010eb,Barnich:2011ty,Barnich:2011mi,Barnich:2013axa,Strominger:2013jfa,Strominger:2014pwa,Flanagan:2015pxa,Barnich:2016lyg,Compere:2018ylh,Nichols:2017rqr,Nichols:2018qac,Bonga:2018gzr,Distler:2018rwu,Ashtekar:2019viz,Ashtekar:2019rpv}. While the asymptotic symmetry group of standard asymptotically flat spacetimes is the BMS group, consisting of Lorentz transformations and supertranslations, the BMS flux-balance laws are associated with the extended BMS group, consisting of both the asymptotic symmetries and the outer symmetries, \emph{i.e.}, the superrotations and superboosts. We derived the global constraints on black hole binary mergers that result from these flux-balance laws by providing the initial and final BMS charges in an arbitrary Lorentz and supertranslation frame. These global constraints can be used by numerical relativists or gravitational wave data analysts as tools to determine the Poincar\'e charge balance as well as the total displacement, spin and center-of-mass memories. We also derived the explicit and exact expansion of all BMS flux-balance laws in terms of the two sets of radiative STF multipoles, which provides consistency constraints on the post-Newtonian/post-Minkowskian formalism and on the radiation-reaction forces of compact binaries. 

Partial radiation gauges are often used to infer the shear resulting from compact binary sources and thereby obtaining the gravitational waveforms. Bondi gauge (or alternatively Newman-Unti gauge) further allows to infer the Bondi mass and Bondi angular momentum aspects which obey evolution laws. In this paper, we fully exploited these evolution laws to derive the exact form of the Poincar\'e flux-balance laws in the radiation zone, independently of the properties of the sources, and independently of the formalism used to study them. Furthermore, we treated comprehensively both the Poincar\'e flux-balance laws and the proper BMS flux-balance laws. We noted in particular that the octupole super-angular momentum flux-balance law arises at the same 2.5PN order as the energy and angular momentum flux-balance laws.

We discussed a two-parameter family of covariant prescriptions for the BMS charges and, in particular, a one-parameter family of covariant prescriptions for the angular momentum, which all lead to vanishing BMS fluxes for non-radiative configurations. We obtained that the prescriptions used in \cite{Thorne:1980ru,Ashtekar:1981bq,Dray:1984rfa,Wald:1999wa,Barnich:2010eb,Barnich:2011mi,Distler:2018rwu,Compere:2018ylh} all agree. We showed that it is the unique prescription within our class that leads to an angular momentum flux that does not admit quadratic terms involving both the parity-odd and parity-even radiative moments. The prescription used in \cite{Pasterski:2015tva,Hawking:2016sgy} instead provides the unique prescription such that the transformation rule of the BMS charges under supertranslations does not involve the shear for non-radiative configurations. We showed that the prescription used to define the center-of-mass in \cite{Blanchet:2018yqa} is not covariant with respect to the metric on the celestial sphere, which implies that this prescribed center-of-mass does not transform covariantly under Lorentz asymptotic symmetries. Instead, we proposed a two-parameter prescription for covariantly defining the center-of-mass, which leads to a new expression for the flux of the center-of-mass to all orders in the radiative multipole expansion. Furthermore, we proposed the supertranslation-invariant definition of Lorentz charges --  the intrinsic Lorentz charges -- for non-radiative configurations, which provides an explicit realization of the dressing procedure described in \cite{Javadinazhed:2018mle} (see also \cite{Moreschi:2002ii,Gallo:2014jda}).

Let us conclude with some future directions. Favata \cite{Favata:2008yd} and Nichols \cite{Nichols:2017rqr,Nichols:2018qac} derived the BMS flux-balance laws using a spherical harmonic basis while we used a basis of symmetric tracefree tensors. The complete comparison of our respective expressions remains to be performed, though for the center-of-mass fluxes they were demonstrated to exactly match. We derived the explicit expressions for the BMS charges for the initial and final states $u \rightarrow \pm \infty$ of black hole mergers in terms of Bondi quantities. A comparison with the geometric expressions derived in \cite{Ashtekar:2019rpv} remains to be performed. We derived the BMS flux-balance laws in terms of radiative multipole moments. The perturbative dictionary between de Donder gauge and Bondi gauge is required in order to rewrite these radiative multipoles in terms of canonical multipoles. The post-Newtonian/post-Minkowskian formalism or, alternatively, the effective field theory approach could then be used to express the canonical multipoles in terms of source parameters and rewrite the BMS flux-balance laws as integro-differential constraints on source parameters. This would allow to fully exploit the infinite-dimensional BMS group to constrain the dynamics of binary systems.

The consequences of the global super-Lorentz flux balance laws and their related spin and center-of-mass memory effects remain to be exploited for numerical simulations of compact binary mergers (see the latest SXS catalog \cite{Boyle:2019kee} which can be analysed using tools defined in Bondi gauge \cite{PhysRevD.88.124010}). The Poincar\'e flux-balance laws allow to deduce the final recoil and angular momentum \cite{Varma:2018aht} or allow to establish the balance of the center-of-mass \cite{Woodford:2019tlo}. The detectability of displacement, spin and center-of-mass memory effects has been partly analyzed but certainly deserves more attention, in particular for space-based gravitational wave observatories. Finally, while there are only three types of memory effects that are relevant for the BMS flux-balance laws, many more persistent gravitational wave observables exist and remain to be classified and analysed for detectability \cite{Flanagan:2018yzh,Compere:2019odm}.

\paragraph{Acknowledgements}
G.C. thanks Glenn Barnich, Nathalie Deruelle, Adrien Fiorucci and Romain Ruzziconi for related discussions. R.O. thanks Simone Speziale for useful conversations.
R.O. and A.S. thank the organizers and participants of the workshop \emph{Precision Gravity: From the LHC to LISA}, especially Miguel Campiglia, Guillaume Faye, Alok Laddha, and Riccardo Sturani for many interesting discussions related to this work. We also thank Luc Blanchet and Rafael Porto for their comments on the manuscript and Luc Blanchet, Guillaume Faye and Abhay Ashtekar for a discussion. We thank Kunal Lobo, Hongji Wei and Samuel Gralla for pointing out typos that were corrected in version 4. This research was partially supported by the Munich Institute for Astro- and Particle Physics (MIAPP) of the DFG cluster of excellence \emph{Origin and Structure of the Universe}.
G.C. is research associate of the F.R.S.-FNRS and he acknowledges support from the IISN convention 4.4503.15 and the COST Action GWverse CA16104.
R.O. is funded by the European Structural and Investment Funds (ESIF) and the Czech Ministry of Education, Youth and Sports (MSMT) (Project CoGraDS - CZ.02.1.01/0.0/0.0/15003/0000437).
A.S. was funded by the F.R.S.-FNRS until September 2019. He currently receives funding from the European Union's Horizon 2020 research and innovation program under the Marie Skłodowska-Curie grant agreement No 801505.

\appendix

\section{Construction of boosted supertranslated Kerr}
\label{app:Kerr}

We will explain here how to arrive to Eq.~\eqref{NAfinal}. We proceed in three steps. First, we first set the Kerr metric in Bondi gauge up to high enough order in the radial expansion: 
\bea
ds^2&=& \left(-1+\frac{2m}{r}-\frac{ma^2(1+3 \cos 2\theta))}{2r^3}+\frac{m^2 a^2 \sin^2\theta}{r^4}+O(r^{-6}) \right)du^2 \nonumber\\
&&-2 \left(1+O(r^{-6})\right)dudr
+\left(\frac{3 m a^2 \sin 2\theta}{r^2}+O(r^{-6})\right)dud\theta \nonumber 
 \eea
 \bea 
&& - \left(\frac{4 m a \sin^2\theta }{r}+\frac{m^2 a^3 (17+23 \cos 2\theta) \sin^2\theta}{2r^4}+O(r^{-5})\right)dud\phi 
 \nonumber \\ 
 &&+ \left(r^2 - \frac{m a^2 \sin^2\theta}{r}+O(r^{-4})\right)d\theta^2  + \left(18 \frac{m^2 a^3 \cos\theta \sin^3\theta}{r^3}+O(r^{-4})\right)d\theta d\phi  \nonumber \\
&& +\left(\sin^2 \theta\,  r^2 + \frac{m a^2 \sin^4 \theta}{r}+O(r^{-4})\right)d\phi^2 . 
\eea
One can convert spherical coordinates $(\theta,\phi)$ to stereographic coordinates $(z,\bar z)$ using $\phi = \frac{i}{2} \log {\frac{\bar z}{z}}$, $\sin\theta = \frac{2 \sqrt{z \bar z}}{1+z\bar z}$.

Second, we acted on this metric with a Lorentz transformation combined with a supertranslation while remaining in Bondi gauge. In order to describe the Lorentz transformation, we denote as $n_i$, $i=1,2,3$, the unit Cartesian vector normal to the sphere, 
\bea
n_i = \left( \begin{array}{c} \sin \theta \cos \phi \\ \sin\theta \sin \phi \\ \cos \theta \end{array} \right) = \frac{1}{1+ z \bar z }\left( \begin{array}{c} z+\bar z \\ i(\bar z - z) \\ z \bar z - 1 \end{array} \right) 
\eea
and $v_i$ an arbitrary boost vector
\bea
v_i =  \left( \begin{array}{c} v_x \\ v_y \\ v_z \end{array} \right) =  \frac{v}{1+ z_s \bar z_s } \left( \begin{array}{c} z_s+\bar z_s \\ i(\bar z_s - z_s) \\ z_s \bar z_s - 1 \end{array} \right) .
\eea

In asymptotically flat spacetimes, a Lorentz transformation acts at leading order in the radial expansion as in Minkowski spacetime. Under a proper Lorentz transformation, the unit normal to the sphere transforms as 
\bea
n_i \rightarrow n_i' = \frac{n_i + \left(-\gamma + c \frac{\gamma - 1}{v^2} \vec{v} \cdot \vec{n}\right)\frac{v_i}{c}}{\gamma \left(1- \frac{\vec{v}}{c} \cdot \vec{n}\right)}+O(r^{-1}). 
\eea
The action on the stereographic coordinates is exactly a $SL(2,\mathbb C)$ transformation, 
\bea
z \rightarrow z' &=& \frac{a z + b}{c z + d}+O(r^{-1}) \equiv G(z) + O(r^{-1}),\qquad ad-bc = 1,\label{sl} \\
\bar z \rightarrow \bar z' &=& \frac{\bar a \bar z + \bar b}{\bar c \bar z + \bar d} + O(r^{-1}) \equiv \bar G(\bar z) + O(r^{-1}) 
\eea
where $\;\bar{}\,$ denotes the complex conjugate. Explicitly,
\bea
a &=& \frac{\gamma (\frac{v}{c}+1) - 1 + z_s \bar z_s (1+\gamma (\frac{v}{c}-1))}{\sqrt{2} \sqrt{\gamma-1}(1+z_s \bar z_s)} ,\\
b&=&\bar c = -\frac{\sqrt{2} z_s \sqrt{\gamma - 1}}{1+ z_s \bar z_s}, \\
d &=&  \frac{\gamma (\frac{v}{c}-1) + 1 + z_s \bar z_s (-1+\gamma (\frac{v}{c}+1))}{\sqrt{2} \sqrt{\gamma-1}(1+z_s \bar z_s)}.
\eea
More generally, a rotation and boost is isomorphic to an arbitrary $SL(2,\mathbb C)$ transformation \eqref{sl}. The resulting angles on the sphere $(\theta',\phi')$ are defined as 
\bea
\phi' \equiv \frac{i}{2}\log {\frac{\bar G(\bar z)}{G(z)}},\qquad \sin \theta' = \frac{2 \sqrt{G(z) \bar G(\bar z)}}{1+G(z) \bar G(\bar z)}. \label{thetap}
\eea
Note the important relationship valid for an arbitrary $SL(2,\mathbb C)$ transformation (rotations do not contribute):
\bea
 \frac{1+G \bar G}{(1+z \bar z)\sqrt{\p_z G \p_{\bar z}\bar G}} = \gamma (1-\frac{\vec{v}}{c} \cdot \vec{n}). \label{rel}
\eea
The leading order Lorentz transformation combined with a supertranslation can then be extended to a 4-dimensional diffeomorphism defined in the radial expansion that enforce Bondi gauge. The subleading components of the diffeomorphism are obtained by solving algebraic constraints that are equivalent to enforcing Bondi gauge. The computation is long but straightforward. 

As a third and final step, we finally read off the resulting Bondi mass and angular momentum aspects and simplify using Eq.~\eqref{rel}. The final result is exactly Eq.~\eqref{mfinal} and Eq.~\eqref{NAfinal}.

\section{Multipole decomposition of the BMS fluxes} \label{App:T1T2}
To compute the right-hand side of Eq.~\eqref{BMSfluxes}, one can use two strategies. The first approach is to rewrite the integrands by expanding the shear tensor $C_{AB}$ according to Eq.~\eqref{Cpm} in terms of the two polarizations  $C^\pm$. Using the identities on the unit sphere of metric $\gamma_{AB}$, $[D_A,D_B]V^C=R^C_{\;\; DAB}V^D$, $R_{ABCD}=\gamma_{AC}\gamma_{BD}-\gamma_{AD}\gamma_{BC}$ and $\eps_{AB}\eps^{CD}=\de_A^C\de_B^D-\de_A^D\de_B^C$, we find
\begin{align}
\dfrac{1}{4}
C_{AB}C^{AB}&=\big(D_A D_B C^+D^A D^B C^+-\dfrac{1}{2}\Delta C^+ \Delta C^+\big)
-2D_A D_B C^+\eps_{C}^{\;\;A}D^B D^C C^- \nonumber \\
& +\big(D_{A} D_B C^-D^A D^B C^--\dfrac{1}{2}\Delta C^- \Delta C^-\big) .\label{energy density}
\end{align}
Note that the  terms quadratic in either $C^-$ or $C^+$ follow the same pattern. The super-angular momentum and super-center-of-mass flux-balance laws \eqref{fluxJ2} and \eqref{sboostlaw} can be expanded similarly. We can then embed the sphere into $\mathbb R^3$ and expand the two functions $C^\pm$ in terms of radiative multipoles according to Eq.~\eqref{dictCp}. 

The second equivalent approach is to embed the sphere into $\mathbb R^3$ from the beginning and to perform the multipolar expansion of the shear tensor  \eqref{CijM} in terms of the radiative multipole moments \eqref{UV}. Both computations are straightforward but lengthy. To write down the result, it is convenient to first define the following three scalar quadratic operators
\begin{subequations}\label{Q}
\begin{align}
\text{Q}^{+}(\bA_L,\bB_{L'}) & \equiv  N_{L-2}N_{L'-2}\bA_{ijL-2}\bB_{ij L'-2} -2 N_{L-1}N_{L'-1}\bA_{i L-1} \bB_{i L'-1}\nonumber\\
&\quad+ \frac{1}{2} N_LN_{L'} \bA_L  \bB_{L'}, \\
\text{Q}^-(\bA_L,\bB_{L'})  & \equiv  \eps_{ijk}n_i \left( N_{L-2} N_{ L'-2}\bA_{jl L-2}\bB_{ k l L'-2}  - N_{L-1} N_{ L'-1}\bA_{j L-1}\bB_{k L'-1} \right),\\
\widehat{\text{Q}}^+(\bA_L,\bB_{L'}) &\equiv N_{L-1}N_{L'-1}\bA_{i L-1}\bB_{i L'-1} - \frac{1}{2} N_L N_{L'} \bA_{L}\bB_{L'}, 
\end{align}
\end{subequations}
the two vector quadratic operators
\begin{subequations} \label{Qoper1}
\begin{align}
\text{Q}^+_k(\bA_L,\bB_{L'})  & \equiv   N_{L-2} N_{L'-3}~{\bA}_{ijL-2}~\bB_{ijkL'-3} - 2 N_{L-1} N_{L'-2}~{\bA}_{iL-1}~ \bB_{ikL'-2}\nonumber\\
&\quad +\dfrac{1}{2} N_{L} N_{L'-1}~{\bA}_{L}~\bB_{kL'-1} ,  \\ 
\widehat{\text{Q}}^+_k(\bA_L,\bB_{L'})  & \equiv  N_{L-2}N_{L'-1} {\bA}_{ki L-2}\bB_{i L'- 1} - N_{L-1}N_{L'} {\bA}_{kL-1}\bB_{L'}\nonumber \\ 
& \quad+\frac{1}{2} N_L N_{L'-1}{\bA}_L \bB_{kL'-1}, 
\end{align}
\end{subequations}
and the two tensor quadratic operators
\begin{subequations} \label{Qoper2}
\begin{align}
\text{Q}^+_{ijk}(\bA_L,\bB_{L'})  &\equiv N_{L-2} N_{L'-3}\bA_{i mL-2}\bB_{jk mL'-3} - N_{L-1} N_{L'-2}\bA_{i L-1}\bB_{jk L'-2}, \\
\text{Q}^-_{ijk}(\bA_L,\bB_{L'})  &\equiv \eps_{ipq}n_p N_{L-2}  N_{L'-1} \bA_{j q L-2} \bB_{kL'-1}.
\end{align}
\end{subequations}
They obey the following properties 
\bea
n_k \text{Q}^+_k = \text{Q}^+, \qquad n_k \widehat{\text{Q}}^+_k \equiv \widehat{\text{Q}}^+ ,\qquad n^i \text{Q}^-_{ijk}=0.
\eea

The quadratic expression appearing in the hard contribution to the supermomentum flux-balance equation reads as
\bea
\!\!\!\!\!\!\dot C^{ij}\dot C_{ij}\!\!\!& \!=\!&\! \!\!\!\sum_{\ell, \ell'=2}^\infty a_\ell a_{\ell'} \Big(\!  \text{Q}^+(\dot \bU_L,\dot \bU_{L'}) + \frac{b_{\ell} b_{\ell'}}{c^2} \text{Q}^+(\dot \bV_L,\dot \bV_{L'}) +\frac{2b_{\ell'} }{c}  {\text{Q}}^-(\dot \bU_L,\dot \bV_{L'})\! \Big).\label{CC}
\eea

The first quadratic operator can then be written as 
\begin{align} \label{T1dec}
\dot C_{ij}\hat \p_k C_{ij} &= P_{kl}\sum_{\ell,\ell'=2}^\infty a_\ell a_{\ell'} \bigg(\text{Q}_l^{(1,1)} (\dot \bU_L,\bU_{L'})+\frac{b_\ell b_{\ell'}}{c^2} \text{Q}_l^{(1,1)}(\dot \bV_L,\bV_{L'}) \nonumber \\
&\qquad \qquad \qquad \qquad +\frac{b_{\ell'}}{c} \text{Q}_l^{(1,2)}(\dot \bU_L,\bV_{L'}) +\frac{ b_{\ell'}}{c}\text{Q}_l^{(1,3)}(\bU_L,\dot \bV_{L'}) \bigg)
\end{align}
where
\begin{subequations}
\begin{align}
\text{Q}_k^{(1,1)}(\dot \bU_L,\bU_{L'})& \equiv (\ell'-2)\text{Q}^+_k(\dot \bU_L,\bU_{L'})-2 \widehat{\text{Q}}^+_k(\dot\bU_L,\bU_{L'}) , \\
\text{Q}_k^{(1,2)}(\dot \bU_L,\bV_{L'})&\equiv  \eps_{kpq}\left[\text{Q}^+_{pqm} (\dot \bU_L,\bV_{L'}) n_m -n_p \text{Q}^+_{iqj}(\dot \bU_L,\bV_{L'}) n_i n_j \right]\nonumber\\
&\quad+ (\ell'-2) \eps_{p ij }n_p \text{Q}^+_{ijk} (\dot \bU_L,\bV_{L'})-\text{Q}^-_{iki} ( \dot \bU_L,\bV_{L'}), \\
\text{Q}_k^{(1,3)}(\bU_L,\dot \bV_{L'})&\equiv \eps_{kpq} n_p \text{Q}^+_{iqj} (\bU_L,\dot \bV_{L'})n_i n_j \nonumber\\
&\quad - (\ell-2) \eps_{pij}n_p \text{Q}^+_{ijk} (\dot \bV_{L'},\bU_{L}) + \text{Q}^-_{iki} (\dot \bV_{L'},\bU_L).
\end{align}
\end{subequations}

The second quadratic operator reads as 
\begin{align} \label{T2dec}
\dot C_{ij}\hat \p_i C_{jk}&= \sum_{\ell,\ell'=2}^\infty  a_\ell a_{\ell'} P_{kl}\bigg(\text{Q}_l^{(2,1)} (\dot \bU_L,\bU_{L'})+\frac{b_\ell b_{\ell'}}{c^2} \text{Q}_l^{(2,1)}(\dot \bV_L,\bV_{L'}) \nonumber \\
&\qquad \qquad \qquad \qquad +\frac{b_{\ell'}}{c} \text{Q}_l^{(2,2)}(\dot \bU_L,\bV_{L'}) +\frac{ b_{\ell'}}{c}\text{Q}_l^{(2,3)}(\bU_L,\dot \bV_{L'}) \bigg)
\end{align}
where
\bea
\text{Q}_k^{(2,1)}(\dot \bU_L,\bU_{L'})&\equiv &(\ell' - 2)\text{Q}^+_k(\dot \bU_L,\bU_{L'})  +\dfrac{\ell' -2}{2}  \widehat{\text{Q}}^+_k(\dot\bU_L,\bU_{L'}),\\
\text{Q}_k^{(2,2)}(\dot \bU_L,\bV_{L'})&\equiv &\frac{1}{2}\text{Q}^-_{kjj} (\bV_{L'},\dot \bU_L) -\frac{1}{2} \text{Q}^-_{iki} (\dot \bU_L,\bV_{L'})-\frac{1}{2}\eps_{kpq}\text{Q}^+_{pqi} (\dot \bU_L,\bV_{L'})n_i \nonumber \\
&& - \frac{\ell' - 2}{2} \eps_{kpq}n_p \text{Q}^+_{q} (\dot \bU_L,\bV_{L'}) + \frac{\ell'-2}{2}\eps_{pij} n_p \text{Q}^+_{ijk} (\dot \bU_L,\bV_{L'})   ,\label{Q22} \\
\text{Q}_k^{(2,3)}(\bU_L,\dot \bV_{L'})&\equiv &-\frac{\ell-2}{4} \eps_{kpq}n_p \text{Q}^+_{iqj} (\bU_L,\dot \bV_{L'}) n_i n_j  -(\ell-2) \eps_{pij } n_p \text{Q}^+_{ijk} (\dot \bV_{L'},\bU_{L}) \nonumber \\
&& - \frac{\ell-2}{4}  \text{Q}^-_{iki} (\dot \bV_{L'},\bU_{L})   .
\eea
Note that $ \text{Q}^+_{iqj} (\bU_L,\dot \bV_{L'}) n_i n_j  =  \text{Q}^+_{qij} (\dot \bV_{L'},\bU_L) n_i n_j $.

The right-hand sides of Eq.~\eqref{BMSfluxes} are combinations of these expressions smeared with BMS symmetry parameters. To perform the integrals of these quantities over the unit sphere, we will use the integrals introduced in Appendix \ref{app:int}.

\section{Integration of tensors on the sphere}
\label{app:int}

We consider the integral 
\begin{align}\label{basic int}
I_L=\oint_S N_{L} 
\end{align}
over the unit sphere $S$, which is fundamental in order to integrate combinations of symmetric trace-free (STF) tensors on the sphere. It can be most easily computed using the generating function $\oint_S e^{-ik\cdot n}=\frac{\sin k}{k}$. We deduce 
\begin{equation}
I_L=\lim\limits_{k\to 0} i^\ell \left(\frac{\pd}{\pd k}\right)_L \left(\frac{\sin k}{k}\right)
\end{equation} 
where $\left(\dfrac{\pd}{\pd k}\right)_{L} \equiv \dfrac{\pd^\ell}{\pd k^{i_1}\cdots \pd k^{i_\ell}}$. Some algebra and combinatorics then gives $I_L=0$ for odd $\ell=|L|$, while for even $\ell$ (see Eq.(2.4) of \cite{Thorne:1980ru}) 
\begin{align}\label{result 1}
\oint_S N_{L} &= \frac{1}{\ell+1}\,  \de_{(i_1i_2}\de_{i_3i_4}\cdots \de_{i_{\ell-1}i_{\ell})}\\ &=\dfrac{1}{(\ell+1)!!}(\de_{i_1i_2}\de_{i_3i_4}\cdots \de_{i_{\ell-1}i_{\ell}}+\text{all ordered permutations})\,,
\end{align}
where an ordered permutation $\{1,\cdots, \ell\}\to \{\sigma_1,\cdots ,\sigma_\ell\}$ is defined as $\sigma_1 =1$, $\sigma_2 \in \{ 2,\dots \ell\}$, $\sigma_3$ smallest integer $\neq \{ 1,\sigma_2\}$, $\sigma_4 \in \{ 2,\dots \ell\} \backslash \{ \sigma_2,\sigma_3\}$, $\sigma_5$ smallest integer $\neq \{ 1,\sigma_2,\sigma_3,\sigma_4 \}$, etc. As an example, $\oint_S N_{ijmn}=\frac{1}{15}(\de_{ij}\de_{mn}+\de_{im}\de_{jn}+\de_{in}\de_{jm})$. Using this fundamental equality, we obtain the following useful formulae. For any given pair of STF tensors $\bA_L$, $\bB_{L'}$, we have
\bea
\bA_L \bB_{L'} \oint_S N_L N_{L'} = \delta_{\ell,\ell'}\, m_\ell \,\bA_L \bB_L,\qquad m_\ell \equiv \frac{\ell !}{(2\ell + 1)!!}\Theta_{\ell}. 
\label{quadint}
\eea
The formula \eqref{quadint} was given in \cite{Thorne:1980ru}. We also have 
\bea \label{quadintbis}
\bA_{iL-1} \bB_{iL'-1} \oint_S N_{L-1} N_{L'-1} = \delta_{\ell,\ell'} m_{\ell-1} \bA_L \bB_L.
\eea
Here, we introduced the discrete step function $\Theta_\ell$ defined as $1$ if $\ell \geq 0$ and 0 if $\ell < 0$. It implements the requirement that the right-hand side of Eq.~\eqref{quadint} is defined only for $\ell \geq0$ and that of Eq.~\eqref{quadintbis} is defined only for $\ell \geq 1$.

For any given triplet of STF tensors $\bA_L$, $\bB_{L'}$ and $\bC_{L''}$, we have 
\bea
\bA_L \bB_{L'} \bC_{L''} \oint_S N_L N_{L'} N_{L''}= \delta_{\ell,\ell',\ell''} m_{\ell,\ell',\ell''} \bA_{L_2 L_3}\bB_{L_1 L_3}\bC_{L_1 L_2}.\label{triple}
\eea
Here, the STF indices are split as $L=L_2 L_3$, $L'=L_1 L_3$, $L''=L_1 L_2$ where $L_{1,2,3}$ are chains of indices with the following ranks
\bea
\ell_1 \equiv \frac{-\ell+\ell'+\ell''}{2},\qquad 
\ell_2 \equiv \frac{\ell-\ell'+\ell''}{2},\qquad
\ell_3 \equiv \frac{\ell+\ell'-\ell''}{2}.\label{defell123}
\eea
The symbol $\delta_{\ell,\ell',\ell''} $ is to ensure that these are integers, \emph{i.e.}, it is defined as
\begin{align}\label{def delta}
    \delta_{\ell,\ell',\ell''}&=\begin{cases}
    1,&  \qquad\ell_1,\ell_2,\ell_3 \in \mathbb Z\\
    0,   &  \qquad          \text{otherwise}
\end{cases}
\end{align}
The integral then amounts to the normalization factor
\bea
m_{\ell,\ell',\ell''} &\equiv  \dfrac{\ell!\ell'!\ell''!}{\ell_1 ! \ell_2 ! \ell_3 !  (\ell+\ell'+\ell''+1)!! }\Theta_{\ell_1} \Theta_{\ell_2}\Theta_{\ell_3} \label{def:m} 
\eea
where $\ell_1,\ell_2,\ell_3$ are functions of $\ell,\ell',\ell''$, as defined in Eq.~\eqref{defell123}. The normalization factor is totally symmetric in its three indices. The step functions $\Theta_\ell$ ensure that $\ell_1$, $\ell_2$, $\ell_3$ are non-negative\footnote{One could include these positivity conditions in the definition \eqref{def delta}, but we need this separation for a nice presentation of our results in Section \ref{sec:flux}.}. This formula was derived in a related form in Eq.~(C2) of \cite{Faye:2014fra}. A closely related quantity $\mathcal C_\ell(0,\ell',m',0,\ell'',m'')$ appears in \cite{Nichols:2017rqr}, itself based on Eq.~(2.20) of \cite{Beyer:2013loa}; see also \cite{Favata:2008yd}. Note that 
\begin{subequations}\label{special case m}
\begin{align}
\delta_{\ell,\ell',0} m_{\ell,\ell',0} &= \delta_{\ell,\ell'} m_\ell,  \\
\delta_{\ell,\ell',1}m_{\ell,\ell',1}  &= m_{\ell}\delta_{\ell,\ell' + 1} + m_{\ell'} \delta_{\ell',\ell + 1} . 
\end{align}
\end{subequations}
Another closely related type of integral that appears for parity odd expressions is the following 
\begin{align}\label{hat m int}
    \eps_{ijk}\int n_i N_{L+1} N_{L'}N_{L''}\bA_{L+1}\bB_{jL'}\bC_{kL''}=&\,\delta_{\ell,\ell',\ell''}\,\widehat m_{\underline{\ell},\ell',\ell''} \,\eps_{ijk}\,\bA_{iL_2L_3}\,\bB_{jL_1L_3}\,\bC_{kL_1L_2}
\end{align}
where $\delta_{\ell,\ell',\ell''}$ determines $\ell_1,\ell_2,\ell_3$ according to Eq.~\eqref{defell123}. The combinatoric function $\widehat{m}$ is defined as 
\begin{align}\label{hat m def}
      \widehat m_{\underline{\,\ell\,},\ell',\ell''}&\equiv\dfrac{\ell+1}{\ell+\ell'+\ell''+3}m_{\ell,\ell',\ell''}\,.
\end{align}
The underlined argument determines the numerator of the coefficient on the right-hand side. Note that due to the presence of $\eps_{ijk}$ in Eq.~\eqref{hat m int}, $n_i$ can only contract with $N_{L+1}$ and thereby leads to Eq.~\eqref{hat m def}. The following relation exists between $m$ and $\widehat m$
\begin{align}
    m_{\ell,\ell',\ell''}&=\widehat m_{\underline{\,\ell\,},\ell',\ell''}+\widehat m_{\ell,\,\underline{\ell'}\,,\ell''}+\widehat m_{\ell,\ell',\,\underline{\ell''}}.
\end{align}
One can also check that 
\begin{align}\label{m hat l''=0}
     \widehat m_{\underline{\ell-1},\ell'-1,0}\;\de_{\ell,\ell'}&=m_\ell \;\de_{\ell,\ell'},\qquad\quad \widehat m_{\ell-1,\ell'-1,\underline{\,0\,}}\;\de_{\ell,\ell'}=\dfrac{m_\ell}{\ell} \;\de_{\ell,\ell'}.
\end{align}

Setting $\ell'' = 1$ and $\bC_i = 1$ in \eqref{triple}, we obtain
\bea
\bA_L \bB_{L'} \oint_S N_L N_{L'} n_i = m_{\ell+1} \delta_{\ell+1,\ell'}\bA_{L} \bB_{iL}+m_{\ell'+1}  \delta_{\ell,\ell'+1}\bA_{iL'} \bB_{L'} .
\label{quadint2}
\eea
Using Eq.~\eqref{quadint} and Eq.~\eqref{quadint2}, the integrals over the sphere of the zeroth and first moment of the two scalar quadratic patterns $\text{Q}^+$ and $ {\text{Q}}^-$ defined in Eq.~\eqref{Q} are given by 
\begin{subequations}\label{EPintegrals}
\begin{align}
\oint_S \text{Q}^+(\bA_L,\bB_{L'}) &= \delta_{\ell,\ell'}  m^+_{\ell} \bA_L \bB_L,  \label{Q1}\\
\oint_S  \text{Q}^-(\bA_L,\bB_{L'}) &=0,\label{Q2}\\ 
\oint_S  \text{Q}^+(\bA_L,\bB_{L'}) n_i &= \delta_{\ell',\ell+1} m^+_{\ell+1} \bA_L \bB_{iL}   +\delta_{\ell,\ell'+1} m^+_{\ell'+1} \bA_{iL'} \bB_{L'}   ,\label{Q3} \\
\oint_S {\text{Q}}^-(\bA_L,\bB_{L'}) n_i &=\delta_{\ell,\ell'} \left(\frac{m_{\ell -1}}{\ell - 1}- \frac{m_{\ell}}{\ell}\right)\eps_{ijk}\bA_{j L-1} \bB_{k L-1}\label{Q4}
\end{align}
\end{subequations}
where we defined 
\bea
m^+_{\ell} \equiv m_{\ell-2}-2 m_{\ell-1}+\frac{1}{2}m_{\ell}.
\eea
Using Eq.~\eqref{quadint2}, we also obtain  the integrals required to compute the angular momentum flux-balance law 
\begin{subequations}
\begin{align}
\oint_S \eps_{ikm} n_m \text{Q}^+_k(\dot \bU_L, \bU_{L'}) &= - \oint_S \eps_{ikm} n_m \text{Q}^+_k(\bU_{L'}, \dot \bU_L) = \delta_{\ell,\ell'}\, m^+_{\ell} \,\eps_{ijk}\, \bU_{j L -1}\dot \bU_{kL-1}, \\
\oint_S \eps_{ikm} n_m \widehat{\text{Q}}^+_k(\dot \bU_L, \bU_{L'}) &= - \oint_S \eps_{ikm} n_m \widehat{\text{Q}}^+_k(\bU_{L'}, \dot \bU_L) \nonumber\\
&= - \delta_{\ell,\ell'} \left(m_{\ell- 1} - \frac{3}{2}m_\ell\right) \, \eps_{ijk} \,\bU_{j L -1}\dot \bU_{kL-1}
\end{align}
\end{subequations}
where it is useful to note
\bea
m_{\ell- 1} - \frac{3}{2}m_\ell = \frac{\ell-1}{\ell+1} m^+_{\ell}.
\eea
We also obtain the integrals required to compute the centre-of-mass flux-balance law 
\begin{subequations}
\begin{align}
\oint_S  P_{ik} \text{Q}^+_k(\dot \bU_L, \bU_{L'}) &= \delta_{\ell+1,\ell'} (m^+_{\ell}-m^+_{\ell+1}) \dot\bU_L \bU_{iL}  - \delta_{\ell,\ell'+1} m^+_{\ell}  \dot\bU_{iL-1} \bU_{L-1} ,\\
\oint_S P_{ik}\widehat{\text{Q}}^+_k(\dot \bU_L, \bU_{L'}) &= -\frac{1}{2}\delta_{\ell+1,\ell'}(m_\ell - m_{\ell+1}) \dot\bU_L \bU_{iL} + \delta_{\ell,\ell'+1} m^+_{\ell}  \dot\bU_{iL-1} \bU_{L-1} .
\end{align}
\end{subequations}

For $\ell'' \geq 2$, the delta function and measure are explicitly given by
\begin{subequations}
\begin{align}
\delta_{\ell,\ell',2} m_{\ell,\ell',2}  &= m_{\ell}\delta_{\ell,\ell' + 2} + \frac{2 \ell}{\ell+1}m_{\ell+1} \delta_{\ell,\ell' }+m_{\ell'}\delta_{\ell+2,\ell'} ,   \\
\delta_{\ell,\ell',3}  m_{\ell,\ell',3} &= m_{\ell}\delta_{\ell,\ell' + 3} + \frac{3 (\ell -1)}{\ell+1}m_{\ell+1}  \delta_{\ell,\ell' + 1}+ \frac{3 (\ell' -1)}{\ell'+1}m_{\ell'+1} \delta_{\ell+ 1,\ell'}+m_{\ell'}\delta_{\ell+3,\ell'} ,\nonumber \\
\vdots\nonumber \\
\delta_{\ell,\ell',\ell''} m_{\ell,\ell',\ell''}&= \sum_{k=0}^{\lfloor \frac{\ell''}{2} \rfloor}m_{\ell,\ell',\ell''}  \delta_{\ell,\ell'+\ell''-2k}\Theta_{\ell'-k}+  \sum_{k=0}^{\lfloor \frac{\ell''-1}{2} \rfloor} m_{\ell,\ell',\ell''} \delta_{\ell',\ell+\ell''-2k}\Theta_{\ell - k}.\label{eqC21}
\end{align}
\end{subequations}
Though the factor of $\Theta_{\ell'-k}$ (respectively $\Theta_{\ell-k}$) is redundant with the term $m_{\ell,\ell',\ell''}$ $\delta_{\ell,\ell'+\ell''-2k}$ (resp. $m_{\ell,\ell',\ell''} \delta_{\ell',\ell+\ell''-2k}$), we want to emphasize that the term is vanishing if $\ell' < k$ (resp. $\ell < k$) as a result of the constraint $\ell_3 \geq 0$. This expression will be used within a double sum $\sum_{\ell,\ell' =2}^\infty$ in the main text. The presence of this discrete theta function will reduce the range of the final sum $\sum_{\ell' = 2}^\infty$ (resp. $\sum_{\ell= 2}^\infty$) to $\sum_{\ell' = \text{max}(2,k)}^\infty$ (resp. $\sum_{\ell = \text{max}(2,k)}^\infty$).

The supermomentum flux-balance law requires the following integrals of $\text{Q}^\pm$ \eqref{Q}:
\begin{subequations}\label{supertranslation STF integrals}
\begin{align}
\oint_S  \text{Q}^+(\bA_L,\bB_{L'}) N_{L''} \bC_{L''} &= \delta_{\ell,\ell',\ell''} m^+_{\ell,\ell',\ell''} \bA_{L_2 L_3}\bB_{L_1 L_3} \bC_{L_1 L_2},  \label{Q1g}\\
\oint_S  \text{Q}^-(\bA_L,\bB_{L'}) N_{L''} \bC_{L''} &= \delta_{\ell-1,\ell'-1,\ell''-1} \left(\widehat m_{\ell-2,\ell'-2,\underline{\ell''-1}}- \widehat m_{\ell-1,\ell'-1,\underline{\ell''-1}}\right)\nonumber \\
& \quad  \times  \eps_{ijk}~\bA_{j L_3 L_2} \bB_{k L_1 L_3} \bC_{i L_1 L_2} \label{Q2g}
\end{align}
\end{subequations}
where we find it useful to define
\begin{subequations}\label{def m^pm}
\begin{align}
m^+_{\ell,\ell',\ell''} &\equiv  m_{\ell-2,\ell'-2,\ell''}-2 m_{\ell-1,\ell'-1,\ell''}+\frac{1}{2}m_{\ell,\ell',\ell''} ,\label{m^+}\\
m^-_{\ell,\ell',\ell''} &\equiv  m_{\ell-1,\ell'-1,\ell''}-\frac{3}{2}m_{\ell,\ell',\ell''} .\label{m^-}
\end{align}
\end{subequations}
From these results, one can recover Eqs.~\eqref{Q1}-\eqref{Q2}-\eqref{Q3}-\eqref{Q4} for $\ell''=0,1$. 

The superboosts and superrotations flux-balance laws require the following integrals: 
\begin{subequations}
\begin{align}
\oint_S  \text{Q}^+_k (\bA_L,\bB_{L'}) N_{L''} \bC_{L''} &= \delta_{\ell,\ell'-1,\ell''} m^+_{\ell,\ell'-1,\ell''}\bA_{L_2 L_3} \bB_{k L_1 L_3}\bC_{L_1 L_2}, \\
\oint_S \widehat{ \text{Q}}^+_k (\bA_L,\bB_{L'})  N_{L''} \bC_{L''} &=  \delta_{\ell,\ell'-1,\ell''}\Big[ \left(m_{\ell-2,\ell'-1,\ell''}-m_{\ell-1,\ell',\ell''}\right) \bA_{k L_2 L_3}\bB_{L_1 L_3}C_{L_1 L_2}\nonumber\\
&\quad \qquad \qquad + \frac{1}{2}m_{\ell,\ell'-1,\ell''} \bA_{L_2 L_3} \bB_{k L_1 L_3} \bC_{L_1 L_2}\Big]. 
\end{align}
\end{subequations}
In parallel with Eq.~\eqref{m^+}, we also define $\widehat{m}^+$ through the function $\widehat{m}$
\begin{align}
    \widehat m^+_{\underline{\ell},\ell',\ell''} &\equiv  \widehat m_{\underline{\ell-2},\ell'-2,\ell''}-2 \widehat m_{\underline{\ell-1},\ell'-1,\ell''}+\frac{1}{2}\widehat m_{\underline{\ell},\ell',\ell''} ,\label{hat m^+}\\
    \widehat m^-_{\underline{\ell},\ell',\ell''} &\equiv \widehat m_{\underline{\ell-1},\ell'-1,\ell''}-\dfrac{3}{2} \widehat m_{\underline{\ell},\ell',\ell''}\label{hat m^-}.
\end{align}

\section{Relating STF tensors to spherical harmonics}
\label{app:sph}

We denote as $\boldsymbol{\alpha}^{\ell m}_L$ as defined in  \cite{Blanchet:2013haa} or $\mathcal Y^{\ell m}_L$ as defined in \cite{Thorne:1980ru} the STF tensors that relate the standard orthonormal basis of spherical harmonics\footnote{They are normalized as $\oint Y_{\ell m} Y^*_{\ell m} = (4\pi)^{-1}$. They do include the $(-1)^m$ Condon-Shortley phase (\emph{e.g.} $Y^{11}=-\sqrt{\frac{3}{8 \pi}}e^{\text{i} \phi}\sin\theta$) which matches with Arfken \cite{garfken67:math}, Thorne \cite{Thorne:1980ru} and Wolfram Mathematica's \textsf{SphericalHarmonicY}.} $Y^{\ell m}$ to the set of STF tensors $\hat {\mathbf N}_L = N_{\langle i_1} \dots N_{i_\ell \rangle}$ (where brackets indicate the STF projection)
\bea
\hat {\mathbf N}_L(\theta,\phi) &=& \sum_{m=-\ell}^\ell \boldsymbol{\alpha}_L^{\ell m}Y^{\ell m}(\theta,\phi) =4\pi m_\ell  \sum_{m=-\ell}^\ell \mathcal Y^{*\ell m}_L  Y^{\ell m}(\theta,\phi), \\
Y^{\ell m}(\theta,\phi) &=& \mathcal Y^{\ell m}_L \hat {\mathbf N}_L(\theta,\phi) = \frac{1}{4\pi m_\ell} \boldsymbol{\alpha}_L^{*\ell m}\hat {\mathbf N}_L(\theta,\phi). 
\eea
We have $\boldsymbol{\alpha}^{*\ell,m}_L=(-1)^m \boldsymbol{\alpha}^{\ell ,-m}_L$. Orthonormality of the spherical harmonics and Eq.~\eqref{quadint} gives
\bea
 \boldsymbol{\alpha}_L^{\ell m}\boldsymbol{\alpha}_L^{*\ell m'} = 4 \pi m_\ell\,  \delta_{m,m'}.
\eea
We define the three vectors transforming under the representation $\bf 3$ of $SO(3)$
\bea
\xi^0 \equiv e_z,\qquad \xi^{\pm 1} \equiv \mp \frac{1}{\sqrt{2}}\left(e_x \pm \text{i} e_y\right). 
\eea
We denote as $\langle \ell'' \ell' m'' m' | \ell m\rangle$ the Clebsch-Gordan coefficients branching the irreducible representations $(\bf{2\ell'+1}) \otimes (\bf{2\ell''+1}) \mapsto (\bf{2\ell+1})$. We have (see (2.26b) of  \cite{Thorne:1980ru}) 
\begin{equation}
  \mathcal Y^{* \ell \, m}_L \mathcal Y^{\ell+1\, m+\mu}_{iL}=\frac{1}{4 \pi m_{\ell+1}}\sqrt{\frac{\ell+1}{2\ell+3}} \langle 1 ~\ell~ \mu~ m | \ell+1~ m+\mu\rangle \xi_i^\mu 
\end{equation}
for $\mu = \pm 1,0$ and 0 otherwise with
\begin{subequations}
\begin{align}
\langle 1 ~\ell~ 0~ m | \ell+1~ m\rangle &= \sqrt{\frac{(\ell -m +1)(\ell + m +1)}{(\ell+1)(2\ell+1)}},\\
\langle 1 ~\ell~ \pm 1~ m | \ell+1~ m \pm 1\rangle &= \sqrt{\frac{(\ell \pm m +1)(\ell \pm m +2)}{2(\ell+1)(2\ell+1)}}.
\end{align}
\end{subequations}

The triple integral of spherical harmonics is 
\bea
\oint Y^{\ell_1 m_1} Y^{\ell_2 m_2} Y^{\ell_3 m_3}= \sqrt{ \frac{2\ell_1+1}{4\pi}\frac{2\ell_2+1}{4\pi}\frac{2\ell_3+1}{4\pi}}\left( \begin{array}{ccc} \ell_1 & \ell_2 & \ell_3 \\ 0&0&0 \end{array}\right)\left( \begin{array}{ccc} \ell_1 & \ell_2 & \ell_3 \\ m_1&m_2&m_3 \end{array}\right)
\eea
where $\left( \begin{array}{ccc} \ell_1 & \ell_2 & \ell_3 \\ m_1&m_2&m_3 \end{array}\right)$ is the Wigner $3j$-symbol. It is the equivalent for spherical harmonics of the triple integral of STF tensors \eqref{triple}.

The radiative mass and current moments in the spherical harmonic basis $\text{U}^{\ell m}$, $\text{V}^{\ell m}$ are related to the STF moments $\bU_L$, $\bV_L$ as \cite{Blanchet:2013haa}
\begin{subequations}
\begin{align}
\text{U}^{\ell m} &= A_\ell ~\boldsymbol{\alpha}_L^{\ell m}~\bU_L, \qquad A_\ell \equiv \frac{4}{\ell!} \sqrt{\frac{(\ell+1)(\ell+2)}{2\ell(\ell-1)}},\\
\text{V}^{\ell m} &= - b_\ell A_\ell ~\boldsymbol{\alpha}_L^{\ell m}~\bV_L,
\end{align}
\end{subequations}
or, conversely, $\bU_L = (A_\ell)^{-1} \mathcal Y^{\ell m}_L \text{U}^{\ell m}$, $\bV_L =- (A_\ell b_\ell)^{-1} \mathcal Y^{\ell m}_L \text{V}^{\ell m}$. 
In the spin weighted decomposition with the conventions of \cite{Blanchet:2013haa},
\bea
h_+ - \text{i} h_\times = \sum_{\ell,m} h^{\ell m}\,  \mbox{}_{-2}Y_{\ell m},\qquad h^{\ell m}= - \frac{G}{\sqrt{2}c^{\ell+2}r}(\text{U}^{\ell m}-\frac{\text{i}}{c}\text{V}^{\ell m})+O(r^{-2}). 
\eea 
The relationship with the Bondi shear is given by \cite{Nichols:2017rqr} (since $h$ in \cite{Nichols:2017rqr} is minus $h$ in \cite{Blanchet:2013haa})
\bea
h_+ -\text{i} h_\times = -\frac{1}{r} C_{AB}\bar m^A \bar m^B+O(r^{-2})
\eea 
where $\bar m^A \p_A = \p_\theta - \text{i}\csc\theta \p_\phi$. 

\bibliography{ref_multipoles}

\end{document}